%% file: paper.tex
\documentclass[letterpaper,10pt,conference]{IEEEtran}
\pdfoutput=1

\newdimen\origiwspc
\newdimen\origiwstr
\usepackage{hyperref}
\usepackage{cite}
\usepackage{stmaryrd}
\SetSymbolFont{stmry}{bold}{U}{stmry}{m}{n}
\usepackage{amssymb}
\usepackage{amsthm}
\usepackage{framed}
\usepackage{tcolorbox}
\usepackage{listings}
\usepackage{diagbox}
\usepackage{bm}
\usepackage{makecell}
\usepackage{multirow}
\usepackage{threeparttable}
\usepackage{todonotes}
\setuptodonotes{fancyline}
\usepackage{mdwlist}
\usepackage{float}
\usepackage{microtype}
\usepackage{mathpartir}
\usepackage{mathrsfs} %
\usepackage{xcolor}
\usetikzlibrary{decorations.pathreplacing, decorations.pathmorphing, shapes.misc} %
\usetikzlibrary{arrows}
\usetikzlibrary{graphs}
\usetikzlibrary{positioning,calc,shapes} 
\DeclareRobustCommand{\thinskip}{\hskip 0.16667em\relax}
\def\emdash{---}
\def\d@sh#1#2{\unskip#1\thinskip#2\thinskip\ignorespaces}
\def\Dash{\d@sh\nobreak\emdash}
\def\Ldash{\d@sh\empty{\hbox{\emdash}\nobreak}}
\def\Rdash{\d@sh\nobreak\emdash}

\usepackage[framemethod=default]{mdframed}
\global\mdfdefinestyle{exampledefault}{%
linecolor=lightgray,linewidth=1pt,%
leftmargin=0cm,rightmargin=0cm
}

\newenvironment{sectionhighlight}{\begin{tcolorbox}[size=small]}{\end{tcolorbox}}

\usepackage{enumitem}
\setlist[itemize]{noitemsep} %
\setlist[enumerate]{noitemsep} %
\setlist[itemize]{leftmargin=*,noitemsep}

\usepackage{tikz}

\definecolor{vertf}{RGB}{0,153,51}
\definecolor{rougef}{RGB}{200,0,0}
\RequirePackage{pifont}
\newcommand{\tick}{\textcolor{vertf}{\ding{51}}}

\newcommand{\osname}{\textsc{Asterios}}
\newcommand{\company}{\textsc{Krono-Safe}}
\newcommand{\educrtos}{EducRTOS}

\newcommand{\multipleinstruction}{\leadsto}

\newcommand{\longversion}[1]{#1}
\renewcommand{\longversion}[1]{}

\newtheorem{definition}{Definition}
\newtheorem{theorem}{Theorem}
\newtheorem{corollary}{Corollary}
\newtheorem{assumption}{Assumption}

\usepackage{thmtools}
\declaretheoremstyle[%
  spaceabove=-6pt,%
  spacebelow=6pt,%
  headfont=\normalfont\itshape,%
  postheadspace=1em,%
  qed=\qedsymbol%
]{mystyle} 
\declaretheorem[name={Proof},style=mystyle,unnumbered,
]{prf}

\definecolor{bluekeywords}{rgb}{0.13, 0.13, 1}
\definecolor{greencomments}{rgb}{0, 0.5, 0}
\definecolor{redstrings}{rgb}{0.9, 0, 0}
\definecolor{graynumbers}{rgb}{0.5, 0.5, 0.5}

\usepackage{amsmath}
\usepackage{gnuplot-lua-tikz}

\author{\IEEEauthorblockN{Olivier Nicole\IEEEauthorrefmark{1}\IEEEauthorrefmark{2},
Matthieu Lemerre\IEEEauthorrefmark{1},
Sébastien Bardin\IEEEauthorrefmark{1} and
Xavier Rival\IEEEauthorrefmark{2}\IEEEauthorrefmark{3}}\\
\IEEEauthorblockA{\IEEEauthorrefmark{1}Université Paris-Saclay, CEA List, Saclay, France} %
\IEEEauthorblockA{\IEEEauthorrefmark{2}Département d’informatique de l’ENS, ENS, CNRS, PSL University, Paris, France}
\IEEEauthorblockA{\IEEEauthorrefmark{3}Inria, Paris, France}\\
olivier.nicole@cea.fr, matthieu.lemerre@cea.fr, sebastien.bardin@cea.fr, xavier.rival@ens.fr}

\date{\today}
\title{No Crash, No Exploit: \\Automated Verification of Embedded Kernels}%

\newcommand{\mysubparagraph}[1]{\smallskip\noindent\textbf{#1}.}

\newcommand{\binseccodex}{\mbox{\textsc{BINSEC/Codex}}}
\newcommand{\binsec}{\textsc{BINSEC}}

\begin{document}

\maketitle

\newcommand{\xxx}[1]{\textcolor{red}{XXX: #1}}
\newcommand{\maybe}[1]{\textcolor{green}{MAYBE: #1}}
\newcommand{\maybenot}[1]{}

%
%

%%%%% Artifact badge %%%%%
\begin{tikzpicture}[remember picture,overlay]
  \node[xshift=-2.75cm,yshift=-5.5cm] at (current page.north east)
    {\href{https://github.com/binsec/rtas2021_artifact}{\includegraphics[width=3cm]{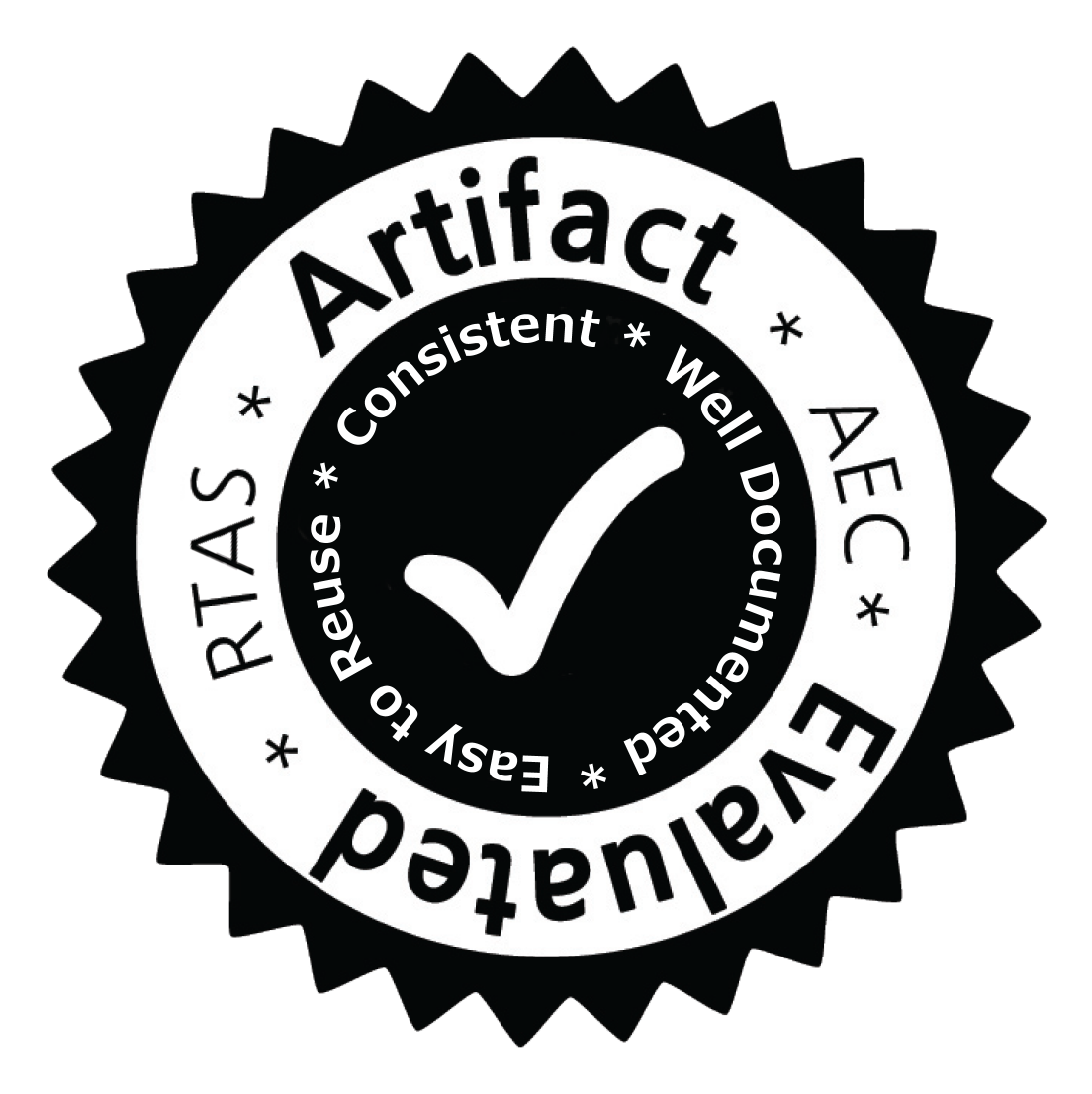}}};
\end{tikzpicture}

\begin{abstract}  The kernel is the most 
  safety- and security-critical component of many computer systems, 
  as the most severe bugs 
  lead to complete system crash or exploit. It is thus desirable to
  guarantee that a kernel is free from these bugs using formal methods,
  but the high cost and expertise required to do so are deterrent to
  wide applicability. 
  We propose a method that can verify both \emph{absence of runtime
    errors} (i.e.\ crashes) and \emph{absence of privilege escalation}
  (i.e.\ exploits) in embedded kernels  from their binary
  executables. %
  The method can verify the kernel runtime independently
  from the application, at the expense of {\it only a few lines} of simple
  annotations. When given a specific application, the method can
  verify simple kernels without any human intervention. 
  We demonstrate our method on two different use cases: 
  we use our tool to help the development of a new embedded real-time kernel, and 
  we verify an existing industrial real-time kernel
  executable with no modification. %
  Results show that the method is fast, simple to use, and can prevent real errors and security vulnerabilities. %
\end{abstract}

\section{Introduction}

The safety and security of many computer systems build  upon that of
its most critical component, the kernel, which provides the
core protection mechanisms. The worst possible defects for a kernel are:
\begin{itemize}
\item \emph{Runtime errors}, which cause the whole system to crash. In
  machine code, there is no undefined behaviour, but faulty execution of an
  instruction in the kernel code that raises an exception
  (e.g.\ illegal opcode, division by zero, or memory protection error)
  is assimilable to a runtime error;
\item \emph{Privilege escalation}, where an attacker bypasses the
  \emph{kernel self-protection} and takes over the whole system. In
  the case of hypervisors, this also corresponds to virtual machine
  escape.
\end{itemize} 

The only way to guarantee that a kernel is free from such defects is
to verify it entirely using formal methods~\cite{klein2009sel4}. %
Manually proving a kernel using a proof assistant~\cite{walker1980specification,bevier1989kit,richards2010modeling,gu2015deep,xu2016practical,klein2009sel4} or deductive verification~\cite{alkassar2010automated,yang2010safe,vasudevan2016uberspark,ferraiuolo2017komodo} can ensure that the kernel
complies with a formal specification, thus reaching a
high level of safety and security. But this is out of
reach of most actors in embedded systems development, because of the
huge effort of writing thousands lines of proof (e.g.\ 200,000 for seL4~\cite{klein2009sel4} or 100,000 for CertiKOS~\cite{gu2016certikos}) and of the
difficulty to find experts in both systems and formal methods. For
such actors, the ideal method would be a fully automated verification
where the system developers only provide their code and, with very little
configuration or none at all, the tool automatically verifies the
properties of interest. In addition, a comprehensive verification
should carry to the binary executable, as 1.~a large
part of embedded kernel code consists in low-level interaction with
the hardware, and 2.~the compilation toolchain (build options,
compiler, assembler, linker) may introduce
bugs~\cite{regehr2011finding}.

Recent so-called ``push-button''
 kernel 
verification methods \cite{dam2013machine,nelson2019scaling,nordholz2020design} are based on symbolic execution \cite{clarke1976symbex,king1976symbex,cadar2013symex}, which
has several advantages: it is sufficiently precise to handle binary
code without getting lost \cite{godefroid2008automated}, and it natively works with logical
formulas and thus can easily be used to prove manually specified
high-level properties. On the other hand,  symbolic execution as
a verification technique suffers from severe limitations, such as the
inability to handle unbounded loops, the need to manually provide all the
kernel invariants (for CertiKOS$^S$~\cite{nelson2019scaling}: 438 lines of
specifications for 1845 instructions, i.e.\ a 23.7\% ratio)  and difficulty to scale because of path explosion. These limitations
are inherent to symbolic execution and cannot be overcome without a
radical change of verification method. It explains, for instance, why 
such  prior  approaches considered only round-robin schedulers \cite{dam2013machine,nelson2019scaling,nordholz2020design}, while   priority-based schedulers (as found in RTOSes) 
working over an arbitrary number of tasks  require an unbounded
loop.

Our goal is to push the boundaries of automated kernel
verification to make it practical for system
developers. We focus on embedded kernels, which are
characterized by mostly static memory allocation, and by the fact that
they typically exist in many variants (depending on the hardware and
application), making the automation of the analysis very important. 
Our contributions are the following:

\begin{itemize}
\item We provide a method (Section~\ref{sec:fully-autom-whole}) for
  \emph{fully automated} verification of \emph{absence of privilege
    escalation} (APE) and \emph{absence of runtime errors} (ARTE) of
  small OS kernels, \emph{in-context} (i.e.\ for a given application layout). We eliminate the need for any manual annotation:
  first by developing a binary-level \emph{abstract
    interpreter}\cite{cousot1977abstract} analyzing the whole \emph{system
  loop} (kernel code + a sound abstraction of user code) to
  automatically \emph{infer} the kernel invariants 
  (rather than merely
  checking them); second by \emph{proving}
  (Theorem~\ref{th:invariant-implies-noescalation}) that APE is an
  \emph{implicit} property, i.e.\ which does not require to write a
  specification (like ARTE~\cite{blanchet2003static});

\item We propose an extension of the method (Section~\ref{sec:param-analys}) for
  \emph{parameterized} kernel verification (i.e.\ verification of the kernel
  independently from the applications). Prior works
  \cite{dam2013machine,nelson2019scaling,nordholz2020design} model
  memory using a \emph{flat model} (the memory is a big
  array of bytes, and addresses are represented numerically) which
  prevents parameterized verification and limits scalability and
  precision \cite{reps2010there} of the verification. We propose a
  type-based representation of memory
  (Section~\ref{sec:type-abstract-domain}) that solves these
  issues. 
  We finally  handle differently  the boot code (whose goal is to establish the system invariant) and the   
  kernel runtime (whose goal is to preserve the system invariant),  
  yielding further precision enhancement (Section~\ref{sec:diff-boot-runt}).   
  If our method requires a tiny number of manual annotations,
  these also replace the \emph{precondition} on the application that
  would be otherwise needed;

\item We conducted a thorough evaluation on two case studies
  (Section~\ref{sec:case-study}), where we demonstrate that 1.~it is
  possible to implement a binary-level abstract interpreter sufficiently precise to
  verify an existing, industrial operating system kernel without
  modification and with only very few annotations ($< 1\%$ ratio); 2.~abstract interpretation is helpful as a continuous
  integration tool to find defects during kernel development,
  especially in embedded kernels that have lots of variants;
  3.~parameterized analysis can scale to large numbers of tasks,
  while our in-context analysis cannot.
\end{itemize}

\section{Problem and motivation}

\begin{figure}[t]
  \lstset{language=C,label= ,caption= ,captionpos=b,numbers=none,morekeywords={int32,int64,type,with,self}}

\begin{lstlisting}
register Int8 sp$'\!$, pc$'\,$, flags$'$, mpu$_1$, mpu$_2$;
Thread *cur; Context *ctx; extern Interface if;
  
kernel(){
  switch( interrupt_number() ){
   case RESET: init(); load_mt(); load_context();
   case YIELD_SYSCALL | TIMER_INTERRUPT: 
    save_context(); schedule(); load_mt(); load_context();
   case ... : ... }
}

save_context(){ ctx->pc = pc$'$; ctx->sp=sp$'$; ctx->flags=flags$'$;}
schedule()    { cur = cur->next; ctx = &cur->ctx;}
init()        { cur = &if.threads[0]; ctx = &cur->ctx;}
load_mt()     { mpu$_1$= &cur->mt->code; mpu$_2$= &cur->mt->data; }
load_context(){ pc$'$= ctx->pc; sp$'$= ctx->sp; flags$'$= ctx->flags; }
\end{lstlisting}

\vspace{-3mm}  
\caption{A simple embedded OS kernel.}%
\label{fig:kernel-example}
\medskip
\begin{lstlisting}[numbers=right]
type Flags = Int8 with (self & PRIVILEGED) == 0
type Context = struct { Int8 pc; Int8 sp; Flags flags; }

type Segment = struct {
  Int8 base; 
  Int8 size_and_rights;
} with self.base $\ge$ kernel_last_addr

type Memory_Table = struct { Segment code; Segment data; }

type Thread = struct { 
  Memory_Table *mt;
  Context ctx;
  Thread *next; }

type Interface = struct { 
   Thread[nb_thread]* threads; 
   (Int8 with self = nb_threads &$\!$& self $\ge$ 1) threads_length; } 

\end{lstlisting}
\vspace{-3mm}
\caption{Types used to interface the kernel and application, with annotations.}
\label{fig:kernel-types}
\end{figure}

We illustrate the problem that our technique solves on  a
simple\footnote{\textls[-1]{While this example is simple to keep it understandable, our method extends to complex embedded kernels featuring e.g.\ shared memory concurrency, complex boot sequence (including dynamic creation of page tables), real-time schedulers, various hardware drivers and system calls (e.g.\ for dynamic thread creation).}}  embedded system, for a fictitious 8-bit hardware
where the user code executes using three registers \textsl{pc}$'$,
\textsl{sp}$'$ and \textsl{flags}$'$, distinct (e.g.\ banked) from
those used for the kernel execution.

\mysubparagraph{Kernel code} The
execution of \emph{user code} (i.e.\ applicative, unprivileged code) 
alternates with that of the \emph{kernel} (which is privileged). For
illustration purposes, we consider the simple kernel with the code in
Fig~\ref{fig:kernel-example} (``\textsl{kernel}'' definition) and the
data structures in Fig~\ref{fig:kernel-types} (the syntax is close to
that of C, and the \textbf{with} annotations can be ignored for now).

The kernel executes whenever an interrupt occurs. If the interrupt
corresponds to a preemption (e.g.\ user code calls a ``yield''
system call, or the kernel receives a timer interrupt), it
saves the values of the registers of the preempted thread
(``\textsl{save\_context}''), determines the next thread to be executed
(``\textsl{schedule}'', here a simple round-robin scheduler), setups
the memory protection to limit the memory accessible by this thread
(``\textsl{load\_mt}''), restores the previous register values of the
thread (``\textsl{load\_context}''), and transitions to user
code. In this example, memory protection is done using two Memory
Protection Unit (MPU) registers (\textsl{mpu}$_1$ or
\textsl{mpu}$_2$): user code can only access the memory addresses allowed by one of the \textsl{mpu} registers, each giving access to one segment (i.e.\ interval of addresses). In addition, unsetting the
\emph{hardware privilege} flag (\textsl{PRIVILEGED} bit in the
\textsl{flags}$'$ register) forbids user code to change the values of
\textsl{mpu}$_1$ and \textsl{mpu}$_2$. 

A special interrupt (RESET) corresponds to the system \emph{boot},
where the kernel initializes the memory
(``\textsl{init}''). 
Additional interrupts (the other \textbf{case}s) would perform
additional actions, like other system calls or interfacing with hardware. 
\begin{figure}[t]
\begin{tikzpicture}[xscale=0.65]

\node[left,align=right] at (11.2,2.6) {\sc kernel\\[-3pt]\sc image};
\draw[dashed] (-2,2.2) -- (11,2.2);
\node[left,align=right] at (11.2,0.5) {\\\sc user\\[-3pt]\sc image};

\newcommand{\hexa}[1]{{\tt {\kern 0.1em}#1}}

\begin{scope}[shift={(0,2.5)}]
\node[left,align=right] at (0,0.25)  {\footnotesize \texttt{\hexa{a0}} \small \textsl{cur}:};
\draw (0,0) rectangle +(1,0.5) node[midway] {\tt \scriptsize \hexa{a7}};

\end{scope}

\begin{scope}[shift={(8,2.5)}]
\node[left] at (0,0.25)  {\footnotesize \texttt{\hexa{a1}} \small \textsl{ctx}:};
\draw (0,0) rectangle +(1,0.5) node[midway] {\tt \scriptsize \hexa{a8}};

\end{scope}

\begin{scope}[shift={(0,1.25)}]
\node[left,align=right,yshift=1.25mm] at (0,0.25)  {\footnotesize \hexa{a2} \small : \\[-3pt] \small Thread[2]};
\draw (0,0) rectangle +(1,0.5) node[midway] {\tt \scriptsize \hexa{ae}};
\draw (1,0) rectangle +(1,0.5) node[midway] {\tt \scriptsize \hexa{c8}};
\draw (2,0) rectangle +(1,0.5) node[midway] {\tt \scriptsize \hexa{d5}};
\draw (3,0) rectangle +(1,0.5) node[midway] {\tt \scriptsize \hexa{01}};
\draw (4,0) rectangle +(1,0.5) node[midway] {\tt \scriptsize \hexa{a7}};
\draw (5,0) rectangle +(1,0.5) node[midway] {\tt \scriptsize \hexa{ae}};
\draw (6,0) rectangle +(1,0.5) node[midway] {\tt \scriptsize \hexa{c8}};
\draw (7,0) rectangle +(1,0.5) node[midway] {\tt \scriptsize \hexa{d8}};
\draw (8,0) rectangle +(1,0.5) node[midway] {\tt \scriptsize \hexa{01}};
\draw (9,0) rectangle +(1,0.5) node[midway] {\tt \scriptsize \hexa{a2}};

\draw[-stealth,thick,shorten >=1pt] (4.5,0.4) node[circle,inner sep=1pt,fill=black] {}  -| (4.5,0.7) -| (4.9,0.5);

\end{scope}

\begin{scope}[shift={(0,0)}]
\node[left,align=right] at (0,0.25)  {\footnotesize \texttt{\hexa{ac}} \small \textsl{if}:\\[-3pt] \small Interface};
\draw (0,0) rectangle +(1,0.5) node[midway] {\tt \scriptsize \hexa{a2}};
\draw (1,0) rectangle +(1,0.5) node[midway] {\tt \scriptsize \hexa{02}};

\end{scope}
  
\begin{scope}[shift={(5,0)}]
\node[left,align=right] at (0,0.25)  {\footnotesize \texttt{\hexa{ae}} \small :\\[-3pt] \small Mem\_Table};
\draw (0,0) rectangle +(1,0.5) node[midway] {\tt \scriptsize \hexa{c0}};
\draw (1,0) rectangle +(1,0.5) node[midway] {\tt \scriptsize \hexa{0f}};
\draw (2,0) rectangle +(1,0.5) node[midway] {\tt \scriptsize \hexa{e0}};
\draw (3,0) rectangle +(1,0.5) node[midway] {\tt \scriptsize \hexa{0f}};

\end{scope}

  \draw[-stealth,thick,shorten >=1pt] (1,2.75) node[circle,inner sep=1pt,fill=black] {}  -| (5.1,1.75);
  \draw[-stealth,thick,shorten >=1pt] (8.5,2.5) node[circle,inner sep=1pt,fill=black] {} |- (7.5,2) -| (6,1.75);
  \draw[-stealth,thick,shorten >=0.25pt] (0.5,0.4) node[circle,inner sep=1pt,fill=black] {} |- (-0.5,0.65) |- (0,1.25);
  \draw[-stealth,thick,shorten >=0.5pt] (9.5,1.35) node[circle,inner sep=1pt,fill=black] {} |- (0,1.00) -| (0,1.25);
  \draw[-stealth,thick,shorten >=1pt] (0.5,1.35) node[circle,inner sep=1pt,fill=black] {} |- (3,0.8) -| (4.9,0.5);
  \draw[-stealth,thick,shorten >=1pt] (5.5,1.35) node[circle,inner sep=1pt,fill=black] {} |- (5.1,0.8) -| (5.1,0.5);

\end{tikzpicture}
\vspace{-5mm}
\caption{Example of a memory dump of the system.}
\label{fig:system-memory-dump}

\end{figure}
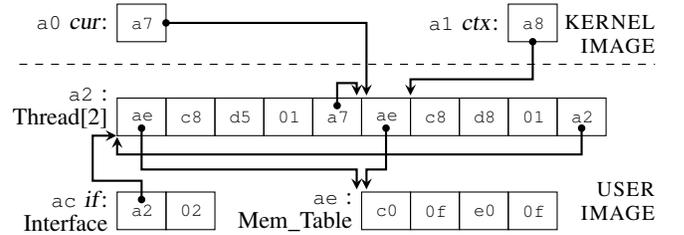

\mysubparagraph{Memory layout and parameterization} Let us now look at
the memory layout of the kernel
(Fig.~\ref{fig:system-memory-dump}). The kernel is
\emph{parameterized}, i.e.\ independent from the \emph{user tasks}
running on it: both the kernel and user tasks can be put in separate
executable \emph{images} and linked at either compile-time or
boot-time. This separation is necessary for closed-source kernel vendors,
and also allows certifying the kernel image independently to reuse this certification in several
applications.

For instance, in Fig.~\ref{fig:system-memory-dump},   addresses
$\mathtt{a0}..\mathtt{a1}$ come from the kernel executable (the \emph{kernel
  image}), while   addresses $\mathtt{a2..b2}$ come from the user tasks
executable (the \emph{user image}). While
Fig.~\ref{fig:system-memory-dump} represents a system composed of two
threads sharing the same memory table, the same kernel image can be
linked to another user image to run a system of 1000 threads, each
with different memory rights.

A consequence of this  parameterization is that the addresses of many
system objects (e.g.\ of type $\mathtt{Thread}$ or
$\mathtt{Memory\_Table}$) vary and are not statically known in the
kernel. It makes the code much harder to analyze  and explains why
in existing automated verification approaches \cite{dam2013machine,nelson2019scaling,nordholz2020design}, these objects must be statically allocated in the kernel, hardcoding a fixed  limit on the number of (e.g.)
$\mathtt{Thread}$s.

\mysubparagraph{Interface and precondition} The kernel and user images
agree on a shared \emph{interface} to work together. In our example,
this interface consists in having the variable
\textsl{if} at an address known by both images; a common
alternative is to use the bootloader to share such information
\cite{multiboot}. 
Also, the system is not expected to work for any user
task image with which the kernel would link: for instance the system
can misbehave, if \textsl{if$\to$threads} points to \textsl{ctx},
to the kernel code or to the stack. Thus, system correctness 
depends on some \emph{precondition} on the provided user image.

\mysubparagraph{Verifications}%
To formally verify that a system satisfies properties such as APE or
ARTE, one has to find \emph{invariants}, i.e.\ properties that are
inductive and initially hold. Experience has
shown that in OS formal verification, {\it ``invariant reasoning
  dominates the proof effort''}\cite{walker1980specification}: e.g.\ in
seL4, 80\% of the effort was spent stating and verifying
invariants~\cite{klein2009sel4}. 
Contrary to existing
manual~\cite{walker1980specification,bevier1989kit,richards2010modeling,gu2015deep,xu2016practical,klein2009sel4},
semi-automated~\cite{alkassar2010automated,yang2010safe,vasudevan2016uberspark,ferraiuolo2017komodo},
or
``push-button''~\cite{dam2013machine,nelson2019scaling,nordholz2020design}
methods, our method automatically infers these invariants. We can perform our verification in two settings:
\begin{itemize}
\item \emph{in-context}, i.e.\ for a given user image layout.
  The invariant that we compute fully
  automatically for the example kernel is given in
  Section~\ref{sec:computed-full-invariant};
\item \emph{parameterized}, i.e.\ independently from the user image. In
  this case, a small number of annotations is needed to describe the
  interface between the kernel and admissible user images; once this is done, the invariant is
  computed fully automatically (for our kernel, it is given in
  Section~\ref{sec:illustr-example-kern-parameterized}).
\end{itemize}

\section{Fully automated in-context analysis}
\label{sec:fully-autom-whole}

We now present our method for static analysis of the kernel
\emph{in-context}, i.e.\ for a given user image. The goal of this
section is to provide a method that  can verify APE and ARTE on small,
simple kernels with no human intervention.

\begin{figure}[t]
\begin{tikzpicture}[xscale=1.1]

  \draw[fill=black!30] (0,1.5) rectangle(8,-1.5);

  \node[below right,scale=0.95,align=left] at (0,1.5) {Whole-system static analysis};

  \node[draw,ellipse,fill=white,align=center,scale=0.9] (s) at (4,0) {kernel\\ exit $\mathcal{E}$};

  \node[draw,ellipse,fill=white,align=center,scale=0.9] (entry) at (6.5,0) {kernel\\ entry};

    \draw (s) edge[red,decorate,decoration={snake,amplitude=.5mm,segment length=2mm,post length=2mm},bend left=20,->, >=latex] node[above,align=center,scale=0.9] {user code} (entry);
    \draw (entry) edge[bend left=20,->, >=latex] node[below,align=center,scale=0.9] {runtime code} (s);        
  
  \node[draw,ellipse,fill=white,align=center,scale=0.9] (s0) at (1.5,0) {initial \\state $s_0$};
  \draw[->] (s0) edge node[above,align=center,scale=0.9] {boot\\code} (s);  

  \node[above right,align=left,scale=0.9] at (0,-1.5) {Finds $\mathcal{I}$ such that: $\ \ \mathcal{I}(s_0)\ \land\ \forall s,s':      \mathcal{I}(s) \land s \to s' \Rightarrow \mathcal{I}(s')$};

\end{tikzpicture}%
\vspace{-5mm}
\caption{Fully-automated in-context analysis on the system loop.}
\label{fig:fully-automated-analysis}
\end{figure}
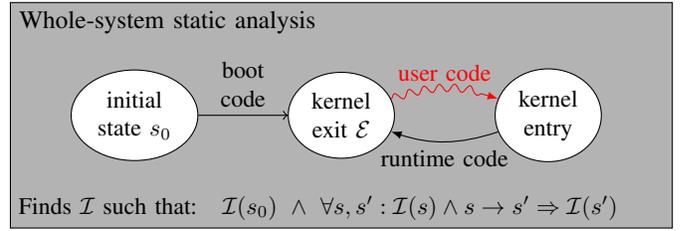

\subsection{Method overview}%
Our method builds upon the following key points:

\mysubparagraph{Key 1: Closed-loop kernel verification} Contrary to
existing methods, we verify the kernel code {\it within}  the whole
\emph{system loop} (Fig~\ref{fig:fully-automated-analysis}) comprising both
the kernel and user code.  Still, we do not need to analyze the user code itself.
Instead, we abstract it by an \emph{empowered}
transition, corresponding to the execution of any
sequence of instructions (Section~\ref{sec:ape-an-implicit}).

\mysubparagraph{Key 2: Binary-level abstract interpretation} We built
a sound static analyzer (Section~\ref{sec:sound-precise-static}) based on {\it abstract interpretation} 
\cite{cousot1977abstract} that works directly on the binary executable
of the kernel. This static analyzer simultaneously reconstructs the
control-flow graph (CFG) of the kernel, and attaches to every
instruction an \emph{abstract state}, summarizing the set of all the
reachable values for each register and memory address. The result of
the analysis is an (inductive) invariant guaranteed to be
correct by construction, i.e.~to encompass all behaviours of the code.  %

\mysubparagraph{Key 3: Kernel implicit properties} We use the computed
invariant to verify whether some \emph{implicit properties}
hold. Implicit properties are properties which do not require writing a manual
specification, and thus can apply to any
kernel. One such property is the well-known absence of run-time errors (ARTE).
More surprisingly,
{absence of privilege escalation} (APE) is also implicit
(Theorem~\ref{th:invariant-implies-noescalation}, Section~\ref{sec:ape-an-implicit}).

\subsection{Illustration on the example kernel}

We illustrate our method on the example kernel of
Fig.~\ref{fig:kernel-example}.

\mysubparagraph{User code transition} The real user-code transition
(which does not appear in the example kernel code) consists in executing the
instruction pointed to by the \textsl{pc} register. The empowered
transition is equivalent to nondeterministic assignment of arbitrary
values to:
\begin{itemize}
\item the user registers \textsl{pc}$'$, \textsl{sp}$'$, and \textsl{flags}$'$;
\item the memory locations that are not protected by either
  \textsl{mpu}$_1$ or \textsl{mpu}$_2$, if \textsl{flags}$'$ has the
  \textsl{PRIVILEGED} bit unset;
\item every memory location, if \textsl{PRIVILEGED} is set in \textsl{flags}$'\!$.
\end{itemize}

Note that the implementation of this transition is not specific to the
kernel that we analyze (it is not ``hardcoded''): it is only specific
to the hardware on which the kernel is implemented.

\mysubparagraph{Computed invariant} \label{sec:computed-full-invariant} Using our binary-level abstract
interpreter on the system composed of the kernel code with the special
user code transition, starting with the initial state of
Fig.~\ref{fig:system-memory-dump}, our analysis computes an inductive
invariant which implies\footnote{\label{note1}The invariant that we actually
  compute is flow-sensitive, and also contain information about the
  kernel code, control-flow graph, stack variables and registers, which we omit for the
  sake of brevity.} that, after boot:%
\begin{itemize}
\item Addresses \texttt{a0} (\textsl{cur}), \texttt{a6} and \texttt{ab} hold either
  \texttt{a2} or \texttt{a7};
\item Address \texttt{a1} (\textsl{ctx} variable) can hold values
  \texttt{a3} or \texttt{a8};
\item Register \textsl{mpu}$_1$ always holds \texttt{ae} and
  \textsl{mpu}$_2$ always holds \texttt{b0};
\item Addresses \texttt{a2} and \texttt{a7} always hold \texttt{ae};
\item Addresses \texttt{a5} and \texttt{aa} and the \texttt{flags}$'$ register can hold any value with
  the \textsl{PRIVILEGED} bit unset;
\item Addresses \texttt{a3}, \texttt{a4}, \texttt{a8}, \texttt{a9},
  and all the other registers can have any value;  
\item Addresses in the range $[\mathtt{c0}..\mathtt{cf}]$ and $[\mathtt{e0}..\mathtt{ef}]$,
  (made accessible by the \textsl{mpu} registers) can have any value;
\item The contents of all the other addresses is constant.
\end{itemize}

\noindent This automatically computed invariant must be manually written in prior 
automated methods \cite{dam2013machine,nelson2019scaling,nordholz2020design}. 

\mysubparagraph{Verifying implicit properties} 
To verify APE, the invariants imply that when the user code executes,
the system is set up such that execution is not \textsl{PRIVILEGED},
and the memory protection tables prevent the user code from modifying
kernel data. Actually, the mere presence of a non-trivial invariant
implies that APE is impossible (details in
Section~\ref{sec:ape-an-implicit}). Verifying ARTE (i.e.\ no faulty execution of an instruction) is simple as our
invariant incorporates the CFG (i.e.\ the set of all instructions
executable by the kernel) and computes a superset of values for every operand of every instruction.%

\subsection{Sound and precise static analysis for binary code}
\label{sec:sound-precise-static}

\mysubparagraph{Abstract interpretation} \emph{Abstract
  interpretation}~\cite{cousot1977abstract,rival2020introduction} is a
general method for building static analyzers that automatically {\it
  infer program invariants} and verify program
  properties. Abstract interpreters are \emph{sound}, in the sense
that the invariants that they infer are correct by
construction. Abstract interpreters have been successfully deployed to
verify large safety- or mission-critical systems from their source
code \cite{blanchet2003static,venet2004precise}.

In essence, abstract interpretation consists in propagating an abstract state that represents a superset of all the possible values for memory and
registers, until a fixpoint is reached. It can be seen as a generalization of
data flow analysis~\cite{kildall1973unified}, with more general domains (e.g.\ intervals, polyhedra) and
widening operators to handle loops~\cite{cousot1977abstract}.

\mysubparagraph{Machine code analysis is challenging}   Binary-level static analysis is particularly challenging  \cite{reps2010there,DBLP:conf/vmcai/BardinHV11,djoudi2016recovering}  
as 1.~the control-flow graph is not known in advance because of computed jumps
(e.g.~\texttt{jmp @sp}) whose resolution requires runtime values,   2.~memory is a single large array of bytes with no prior typing nor partitioning,  and 
 3.~data manipulations are very low-level (masks, flags, legitimate overflows,
 low-level comparisons, etc.).  In other words, it is prone to imprecisions, and
 imprecisions can often snowball, up to a point that the analyzer can only
 return the set of all states (i.e.\ the trivial invariant, denoted $\top$).

\mysubparagraph{Our precise binary-level static analysis}
\label{sec:static-analysis}
\label{sec:static-analysis-machine-code}
To have sufficient precision, our analyzer builds upon state-of-the-art
techniques for machine code analysis. The full formal description of
our static analyzer is out of the scope of the paper, but is available in a
technical report~\cite{nicole2021binseccodex}. We summarize here
the most important techniques that we use:

\begin{itemize}[leftmargin=0.6em]

  \item \emph{Control flow:} Control and data are strongly interwoven at binary level, since  
  resolving computed jumps requires to precisely track  
  values in registers and memory.
  Following Kinder~\cite{kinder2009abstract} or V\'edrine et al.~\cite{DBLP:conf/vmcai/BardinHV11}, 
  our analysis computes simultaneously the CFG and value abstractions;

 \item \emph{Values:}  We mainly use efficient non-relational abstract domains
  (reduced product of the signed and unsigned meaning of
  bitvectors~\cite{djoudi2016recovering},  congruence
  information~\cite{granger1989congruences}), complemented with
   symbolic relational information
  \cite{mine2006symbolic,gange2016uninterpreted,djoudi2016recovering} for local simplifications 
  of sequences of machine code;

 \item \emph{Memory:} Our memory model is ultimately byte-level in order to deal with 
  very low-level coding aspects of kernels. Yet, 
  as representing each memory byte  separately is inefficient and
  imprecise,  we use a stratified representation of memory caching 
  multi-byte loads and stores, like Min\'e~\cite{mine2006field}.  Moreover,
  we do not track memory addresses whose contents is unknown;  

\item \emph{Precision:} To
  have sufficient precision, notably to enable strong updates~\cite{rival2020introduction}  to the
  stack, our analysis is flow-sensitive and fully context-sensitive (i.e.\
  inlines
  the analysis of functions), which is made possible by the small size
  and absence of recursion  typical of microkernels. Moreover, we unroll
  the loops when the analysis finds a bound on their number of iterations;

\item \emph{Concurrency:}  We handle  shared memory zones 
  through   a flow-insensitive abstraction  making them
  independent from  thread
  interleaving~\cite{venet2004precise}. Our weak shape abstract domain
  (Section~\ref{sec:type-abstract-domain}) represents one part of the
  memory in a flow-insensitive way. For the other zones we identify
  the shared memory zones by intersecting  the addresses read and
  written by each thread \cite{mine2011static}, and only perform weak
  updates on them.

\end{itemize}

\subsection{APE is an implicit property}
\label{sec:ape-an-implicit}

We will show that the
mere existence of an invariant implies absence of privilege escalation.

We model the execution of code as a transition system
$\langle \mathbb{S}, \mathbb{S}_0, {\to} \rangle$, where $\mathbb{S}$ is the set
of all
states, $\mathbb{S}_0$ the set of initial states, and $\to$
corresponds to the transition from one instruction to the next. 

\begin{definition}[Privilege escalation]
  We define \emph{privilege escalation} of a transition system
  $\langle \mathbb{S}, \mathbb{S}_0, {\to} \rangle$ as reaching a state which is
  both \emph{privileged} and \emph{not kernel-controlled}.
\end{definition}

Thus, an attacker can escalate its privilege by either gaining control
over privileged kernel code (e.g., by code injection), or by leading
the kernel into giving it its privilege (e.g., by corrupting the
\textsl{flags}$'$ register).

The model that we analyse is
$\langle \mathbb{S}, \mathbb{S}_0, {\multipleinstruction}\rangle$, where we have
replaced the normal transition $\to$ (corresponding to the execution of
the next instruction) by the \emph{empowered} transition
${\multipleinstruction}$ (which non-deterministically executes any sequence of
instructions permitted by the hardware). We call this transition
\emph{empowered} since it gives more power to the user to attack the
kernel, as formalized in the following theorems:

\begin{theorem}
  The set of reachable states for the
  $\langle \mathbb{S}, \mathbb{S}_0, {\to} \rangle$ transition system
  is included in the set of states reachable for
  $\langle \mathbb{S}, \mathbb{S}_0, {\multipleinstruction}
  \rangle$.
\end{theorem}
\begin{prf}
  This follows directly from the fact that for every $s$,
  $\{s': s \to s' \} \subseteq \{ s': s \multipleinstruction s' \}$
\end{prf}

\begin{corollary} \label{th:privilege-escalation-new-transition-system}
  If there are no privilege escalation in the transition system
  $\langle \mathbb{S}, \mathbb{S}_0, {\multipleinstruction}\rangle$,
  there are also no privilege escalation in the transition system
  $\langle \mathbb{S}, \mathbb{S}_0, {\to}\rangle$
\end{corollary}
\begin{prf}  
  This is the contrapositive of the fact that if a state exists in
  $\langle \mathbb{S}, \mathbb{S}_0, {\to} \rangle$ where privilege
  escalation happens, this state also exists in
  $\langle \mathbb{S}, \mathbb{S}_0, {\multipleinstruction}\rangle$.
\end{prf}

Thus, verifying APE on a
$\langle \mathbb{S}, \mathbb{S}_0, {\multipleinstruction}\rangle$ system implies
APE on the real $\langle \mathbb{S}, \mathbb{S}_0, {\to}\rangle$ system.  

Now, to turn APE into an implicit property, we rely on the following
assumption:

\begin{assumption} \label{th:assumption}
  Running an arbitrary sequence of privileged instructions allows to
  reach any state of $\mathbb{S}$.
\end{assumption}

This assumption is very reasonable, as privileged instructions can do
anything that can be done by software. Using it, we can prove the
following theorems:

\begin{theorem} 
  If a transition system
  $\langle \mathbb{S}, \mathbb{S}_0, {\multipleinstruction} \rangle$
  is vulnerable to privilege escalation, then the only satisfiable state property
  in the system is the \emph{trivial state property $\top$}, true for every state.
  \label{th:priv-escal-state-invariant}
\end{theorem}
\begin{prf}
  If a privilege escalation vulnerability exists, then any state can be
  reached (Assumption~\ref{th:assumption}). Hence, the only state
  invariant of the system %
  is~$\top$.
\end{prf}%
\noindent We define a \emph{non-trivial state invariant} as  any state invariant different from~$\top$.

\longversion{Note that on the contrary, if any state can be reached,
  then obviously a privileged attacker-controlled state can be
  reached, provided that such a state exists (if it does not, then
  verifying  absence of privilege escalation is pointless).}

\begin{theorem}
  If a transition system satisfies a non-trivial state invariant, then it is
  invulnerable to privilege escalation attacks.%
  \label{th:invariant-implies-noescalation}
\end{theorem}
\begin{prf} By contraposition, and the fact that state invariants
  are valid state properties. 
\end{prf}

Theorems \ref{th:priv-escal-state-invariant}  \& \ref{th:invariant-implies-noescalation}  have two crucial practical implications:
\begin{itemize}
\item If privilege escalation is possible, \emph{the only state property that holds in the system is $\top$}, making it impossible to prove definitively
  any other property. Thus, \emph{proving  absence of
    privilege escalation is a necessary first step} for any formal
  verification of an OS kernel; 

\item The proof of \emph{any} state property different from $\top$ implies as a
  byproduct the existence of a piece of code able to protect itself
  from the attacker, i.e.\ a kernel with protected privileges. In
  particular, we can prove absence of privilege escalation automatically, by successfully inferring \emph{any} non-trivial state invariant with a sound static analyzer.
\end{itemize}

\section{Parameterized analysis}
\label{sec:param-analys}

\subsection{Shortcomings of the flat memory model}

The fully automated verification works in practice for small, simple kernels,
but has some shortcomings. The root of the problem lies in the
\emph{flat memory model} (also used by all the previous binary-level
automated methods
\cite{dam2013machine,nelson2019scaling,nordholz2020design}),
i.e.\ viewing the memory as a single large array of bytes. This model
poses three distinct problems:
\begin{itemize}

\item \emph{Robustness}, because imprecisions can cascade so much that
  the analysis can no longer compute an invariant. This would happen
  for instance if, in our example kernel, the set of possible values for \textsl{cur},
  $\{\mathtt{a2},\mathtt{a7}\}$, was over-approximated by the interval
  $[\mathtt{a2}..\mathtt{a7}]$;

\item \emph{Scalability}, because of the need to enumerate large
  sets. For instance, on a system with 1000 tasks, the values for
  \textsl{mpu}$_1$ may point to 1000 different arbitrary addresses,
  which must be enumerated if we want to be robust;

\item \emph{Over-specialization}: it is impossible to analyze programs
  without knowing the precise memory layout of all the data. In
  particular, the example kernel cannot be analyzed independently from
  the user image, even if this kernel is independent from the user
  image.

\end{itemize}

\noindent We propose an abstraction that lifts those three limitations,
at the expense of  writing a small number of simple annotations.
  
\subsection{Method overview}

Our method extends that of Section~\ref{sec:fully-autom-whole} with
the following key points:

\mysubparagraph{Key 4: a type-based shape abstract domain} Instead of
representing the memory and addresses numerically, we reason
(Section~\ref{sec:type-system-weak}) about the \emph{shape} of the
(user image) memory based on the \emph{types} used to access
it. Specifically, we verify that the kernel preserves the 
\emph{well-typedness} of the user image (according to provided types,
i.e.\ Fig.~\ref{fig:kernel-types} for the example kernel), and we use
the types to compute the invariant over the values handled by the
kernel. This solves the robustness and scalability issues of the flat
model and allows \emph{parameterized analysis}, i.e.\ to verify the
kernel independently from the  user image.
The kernel memory is still represented numerically, which allows to
verify the kernel using its \emph{raw binary}.

\mysubparagraph{Key 5: differentiated handling of boot and runtime
  code} We propose to handle kernel runtime and boot
differently. %
On the one hand, the kernel runtime \emph{preserves} existing data
structure invariants, while the boot code \emph{establishes} them by
initializing the structures. This makes the runtime suitable for
verification by our shape domain, which aims at verifying the
preservation of memory invariants. On the other hand, when the user
image is known, boot code execution is mostly deterministic (only
small sources of non-determinism remain: multicore handling, device
initialization, etc.); this makes it easy to analyze it robustly using
fully automated in-context analysis
(Section~\ref{sec:fully-autom-whole}). Based on these observations we
propose a method (Section~\ref{sec:diff-boot-runt}) where boot code is
verified using the fully automated in-context analysis, and the
runtime code is verified using the parameterized static analysis.

\mysubparagraph{Summary: The 3-step method}\label{sec:3-step-method} We propose a 3-step method that relies on these keys: 
\begin{itemize}
\item {\emph{Step 1}:} \emph{Lightweight annotation of the interface
    types}.  These types (Fig~\ref{fig:kernel-types}) are known as
  they are needed to develop applications using the kernel, yet additional
  annotations are necessary (Section~\ref{sec:type-system-weak}).

\item {\emph{Step 2}:} \emph{Parameterized static analysis of the
    kernel}. The result of this step is an inductive property
  $\overline{\mathcal{I}}$, which is an invariant from any state where
  this property is true.
\item {\emph{Step 3}:} \emph{Base case checking} consists in verifying
  that the parameterized inductive invariant initially holds.

\end{itemize}

\subsection{Illustration on the example kernel} \label{sec:illustr-example-kern-parameterized}
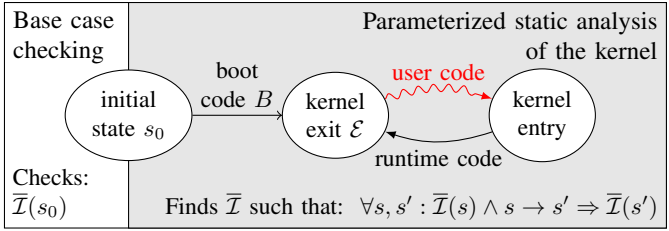
\begin{figure}[t]

  \begin{tikzpicture}[xscale=1.1]
  \draw[fill=white] (0,1.5) rectangle (8,-1.5);
  \draw[fill=black!10] (1.5,1.5) rectangle(8,-1.5);

  \node[below left,scale=0.95,align=right] at (8,1.5) {Parameterized static analysis\\ of the kernel};
  \node[below right,scale=0.95,align=left] at (0,1.5) {Base case\\checking};

  \node[above right,scale=0.9,align=left] at (0,-1.5) {Checks: \\ $\overline{\mathcal{I}}(s_0)$};
  
  \node[draw,ellipse,fill=white,align=center,scale=0.9] (s) at (4,0) {kernel\\ exit $\mathcal{E}$};

  \node[draw,ellipse,fill=white,align=center,scale=0.9] (entry) at (6.5,0) {kernel\\ entry};

    \draw (s) edge[red,decorate,decoration={snake,amplitude=.5mm,segment length=2mm,post length=2mm},bend left=20,->, >=latex] node[above,align=center,scale=0.9] {user code} (entry);
    \draw (entry) edge[bend left=20,->, >=latex] node[below,align=center,scale=0.9] {runtime code} (s); 
  
  \node[draw,ellipse,fill=white,align=center,scale=0.9] (s0) at (1.5,0) {initial \\state $s_0$};
  \draw[->] (s0) edge node[above,align=center,scale=0.9] {boot\\ code $B$} (s);  
  
  \node[above left,scale=0.9] at (8,-1.5) {Finds $\overline{\mathcal{I}}$ such that: $\ \forall s,s':      \overline{\mathcal{I}}(s) \land s \stackrel{}{\to} s' \Rightarrow \overline{\mathcal{I}}(s')$};

\end{tikzpicture}%
\vspace{-5mm}
\caption{Parameterized verification when user memory starts initialized.}
\label{fig:complete-verification-simple-case}
\end{figure}

We illustrate how
the 3-step method is applied on the example kernel
(Fig.~\ref{fig:complete-verification-simple-case}).

\label{sec:illustr-example-kern}

\mysubparagraph{1. Lightweight type annotation} First, we annotate the
types accessible from \textsl{Interface} with annotations.  In larger
kernels, this mostly consists in indicating the size of arrays
(Fig.~\ref{fig:kernel-types}, line 17), and which pointers may
be null. We also specify
that the memory zone accessible to a task is disjoint from the kernel address
space (Fig.~\ref{fig:kernel-types}, line 7), and that saved user flags
are unprivileged (Fig.~\ref{fig:kernel-types}, line 1). Note that
these manual annotations are smaller (and simpler) than the full
invariant (Section~\ref{sec:computed-full-invariant}); this is even
more true on larger kernels.

Concretely, what we call annotations consists in a set of types like in
Fig.~\ref{fig:kernel-types}. These types must be written in a separate file and
provided to our tool along with the kernel executable. No binary or source code
is annotated; instead, our tool only requires to know the address and the type
of the interface entry point (type $\mathtt{Interface}$ at address $\mathtt{ac}$ in the example), i.e.\
the memory location that the kernel should use to access any element of the
interface. The set of types is necessarily provided by the kernel (and bootloader) developers as
part of the software development kit; the ``\textbf{with}'' predicates required to complete the
verification can also be provided by the documentation, by kernel developers, or finding them
can be part of the verification work. We give practical details about this in
the experimental section (Section~\ref{sec:case-study}).

\mysubparagraph{2. Parameterized static analysis of the kernel} Assuming that
these annotations describe the user image, and that the variable
\textsl{if} has type
\lstinline{Interface *}, the analysis computes (using the kernel image) a parameterized invariant $\overline{\mathcal{I}}$ (Fig.~\ref{fig:complete-verification-simple-case}) implying\textsuperscript{\ref{note1}} the following invariant (after boot):

\begin{itemize}
\item Address \texttt{a0} (\textsl{cur} variable) has an admissible $\mathtt{Thread *}$ value;
\item Address \texttt{a1} (\textsl{ctx} variable) has an admissible $\mathtt{Context *}$ value;
\item Registers \textsl{mpu}$_1$ and \textsl{mpu}$_2$ have admissible $\mathtt{Segment}$ value;
\item Register \textsl{flags'} has an admissible $\mathtt{Flags}$ value;  
\item The memory in the user image is well-typed, i.e.\ matches the
  types of Fig.~\ref{fig:kernel-types}.
\end{itemize}

We postpone the exact meaning of the invariant, including definition of ``well-typed''  and ``admissible values'' for a type, to
Section~\ref{sec:type-system-weak}. For the user image of
Fig.~\ref{fig:system-memory-dump}, the above invariant corresponds to
that of Section~\ref{sec:computed-full-invariant}: for instance, the set of admissible values for the type \lstinline{Thread*} in that case is $\{\mathtt{a2}, \mathtt{a7} \}$. However, note that instead of describing the
invariant for one given user image, this new invariant describes what
happens for \emph{all} (well-typed) user images, hence is parameterized. 

\mysubparagraph{3. Base case checking} It remains to be checked whether our
assumption on the initial state ($\overline{\mathcal{I}}(s_0)$) holds,
i.e.\ to verify that, a given user image is well-typed (i.e.\ matches the annotated types),
and that \textsl{if} is the address of an
\lstinline{Interface *}. The precise description of this process (a
kind of type checking) is given in Section~\ref{sec:type-system-weak}; it consists in
particular in checking that the value contained in memory locations supposed to
hold a type $t$, is indeed admissible for $t$; e.g.\ that address $\mathtt{a6}$, of type \lstinline{Thread *}, holds either $\mathtt{a2}$ or $\mathtt{a7}$.

\medskip

Note that the technique is no longer fully automated, as the user has
to provide type annotations (even if most of it comes from the
existing types of the interface). But let us observe that the example
kernel works correctly only when linked with a well-typed user image
(according to Fig.~\ref{fig:kernel-types}), and thus \emph{parameterized
  verification of this kernel is impossible} without limiting, using a
manual \emph{precondition}, the admissible user image. Thus in
general, \emph{parameterized kernel verification is impossible without
  user-supplied annotations}, given here using the type annotations.

\subsection{Differentiating boot and runtime code}
\label{sec:diff-boot-runt}

In the example system, all the data structures in the user image
are already initialized and thus are initially well-typed according to
Fig.~\ref{fig:kernel-types}. In other systems this might not be the
case: for instance, the values of the \textsl{next} field in
\textsl{Thread}s could be uninitialized, and the boot code would have
to create the circular list; or the \lstinline{Memory_Table} table
could be dynamically allocated and filled during the boot, as in our case studies.

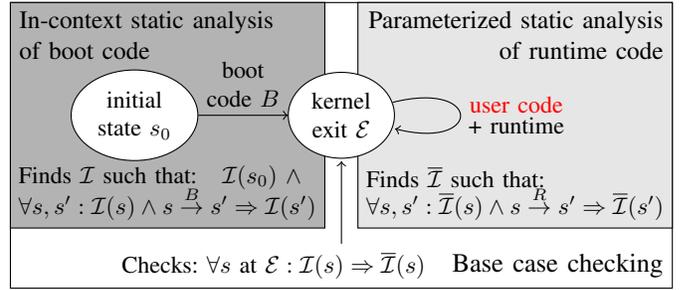
\begin{figure}[t]
\begin{tikzpicture}[xscale=1.1]

  \draw[fill=white] (0,1.5) rectangle (8,-2.3);  
  \draw[fill=black!30] (0,1.5) rectangle(3.8,-1.5);
  \draw[fill=black!10] (4.2,1.5) rectangle(8,-1.5);

  \node[below right,scale=0.95,align=left] at (0,1.5) {In-context static analysis\\ of boot code};
  \node[below left,scale=0.95,align=right] at (8,1.5) {Parameterized static analysis\\ of runtime code};
  \node[left,scale=1,align=right] at (8,-2) {Base case checking};    

  \node[draw,ellipse,fill=white,align=center,scale=0.9] (s) at (4,0) {kernel\\ exit $\mathcal{E}$};
  \draw[->] (s) edge[loop right] node[right,align=center,scale=0.9] {\textcolor{red}{user code}\\[-3pt]+ runtime} (s);

  \node[draw,ellipse,fill=white,align=center,scale=0.9] (s0) at (1.5,0) {initial \\state $s_0$};
  \draw[->] (s0) edge node[above,align=center,scale=0.9] {boot\\ code $B$} (s);  
  
  \node[above left,scale=0.9,align=left] at (8,-1.5) {Finds $\overline{\mathcal{I}}$ such that: \\[-1pt] $\forall s,s':      \overline{\mathcal{I}}(s) \land s \stackrel{R}{\to} s' \Rightarrow \overline{\mathcal{I}}(s')$};

  \node[above right,align=left,scale=0.9] at (0,-1.5) {Finds $\mathcal{I}$ such that: $\ \ \mathcal{I}(s_0)\ \land$\\[-1pt] $\forall s,s':      \mathcal{I}(s) \land s \stackrel{B}{\to} s' \Rightarrow \mathcal{I}(s')$};

  \node[align=left,scale=0.9] (label) at (4,-2) {Checks: $\forall s$ at $\mathcal{E}: \mathcal{I}(s) \Rightarrow \overline{\mathcal{I}}(s)$\hspace{2cm}};
  \draw (label) edge[shorten >=1pt,->] (s);
  
\end{tikzpicture}%
\vspace{-5mm}
\caption{Parameterized verification when user memory needs initialization.}
\label{fig:complete-verification-differentiated-boot}
\end{figure}

To handle these cases, we propose to perform the base case checking
when the kernel has finished booting and enters its main loop
(Fig.~\ref{fig:complete-verification-differentiated-boot}), rather
than at the beginning of the execution. For this, we:
\begin{itemize}
\item Perform the parameterized analysis of the kernel starting from
  an initial state where the user image would already be initialized.
  Even if the true initial state is not initialized, we still get a
  parameterized invariant $\overline{\mathcal{I}}$ on the
  kernel runtime which is inductive by construction;

\item Perform a fully-automated in-context analysis of the kernel
  boot code with the given user image to get an invariant
  $\mathcal{I}$ at the end of the boot code;
  
\item Check that all the ``in-context'' states at the end of the
  boot match the parameterized invariant, i.e.\ check that
  $\forall s,\ \mathcal{I}(s)\Rightarrow\overline{\mathcal{I}}(s)$.

\end{itemize}

Once this is done, the full system is verified: the combination of
both analyses imply the existence of an invariant, and thus APE;
moreover, both analyses allow to check that there is no software
exception neither in the boot nor in the runtime.

This method uses two characteristics of the boot code:
\begin{itemize}
\item First, one goal of the boot code is to
  initialize the data structures so that they are well-typed according to
  Fig.~\ref{fig:kernel-types}. The initial value for these types is
  unimportant as they will be overwritten, so the code will also work
  if these data structures are initialized.
\item The execution of the boot code is almost deterministic (except
  for hardware device handling and multicore execution) once the user
  image is known, thus the fully-automated in-context static analysis
  on this case is both robust and scalable (as it propagates mostly
  singleton values).
\end{itemize}  

\noindent Note that this method avoids the need to specify (and annotate) the
code with the actual precondition on the user image; here, the
precondition that we verify is that ``the user image should be such
that its initialization will make it well-typed'', using the fact that
it is easier to describe the runtime types than the initial types.

\subsection{The type system of the weak shape domain}%
\label{sec:type-system-weak}\label{sec:type-abstract-domain}

Our type-based abstract domain is designed to verify the preservation
of the memory layout of the interface, expressed using a particular
dependent type system. %
A full formal presentation of the domain would be out of the scope of
the paper, but here we introduce the type system and how
well-typedness enforces the preservation of memory layouts. Following our example and for the sake of simplification, all scalar types have size 8 bits, but in practice our system handles multi-byte words.

\begin{figure}
% \todo[inline]{J'ai un peu corrigé, mais ca ne va toujours pas: double utilisation de n, et la distinction
%   entry types et cell types est pas claire, il manque la grammaire des prédicats...,
%   symbolic constants n'est pas dans le texte...}  
% \begin{align*}

     \hspace{-2pt}%
     $\begin{array}{rrll}
      \mathbb{T} \ni t,u &::=& \mathtt{Word} & \textrm{any word}\\
              &|& \mathtt{Int8}  &\textrm{non-pointer} \\
              &|& n              & \textrm{type name} \\
                     &|& {t\hspace{0.2pt}}_i{*}          & \textrm{pointer} \\
%              &|& \omit\rlap{$\ \,\textbf{struct}\ \left\{ \right. {(t\ \langle\mathit{field}\rangle\textbf{;})}\texttt{*} \left. \right\}$} & \textrm{record} \\        
              &|& \textbf{struct}\ \left\{ \right. {(t\ \langle\mathit{field}\rangle;)}^* \left. \right\} & \textrm{record} \\
              &|& t[e] & \textrm{array with size} \\
                     &|& t\ \textbf{with}\ p & \textrm{refinement type} \\
      \mathbb{T}\!\!\times\!\!\mathbb{N}\! \ni\! t_i &::=& t_i & \textrm{type with offset} \\
        \\
        \mathcal{N} \ni n &::=& \omit\rlap{$\ \,\mathtt{Flags} \ |\ \mathtt{Context} \ |\  \ldots$}  & \textrm{type name} \\
        e &::=& i \in \mathbb{N}          &  \textrm{numeric constant}  \\
          &|& \mathtt{nb\_threads}\ |\ \ldots & \textrm{symbolic constant}  \\
          &|& e + e \ |\ \ldots              &  \textrm{binary op} \\
%      c &::=& \omit\rlap{$\mathbb{N} \cup \{\mathtt{nb\_threads}, \ldots\}$}  & \textrm{(symbolic) constant} \\
        p &::=& \textbf{self} \le e \,|\,\textbf{self}\&e = e \,|\,\ldots & \textrm{predicate}
     \end{array}$
%    & \\
%    & \mathcal{C} = \{\mathtt{nb\_threads},\ldots\}\text{ is the set of} \\
%    & \text{symbolic constants}
%  \end{align*}%
  \caption{(Simplified) grammar of types.}%
  \label{fig:grammar-types}
\end{figure}

\mysubparagraph{Annotated types}  Fig.~\ref{fig:kernel-types} consists in a
sequence of \emph{type definitions}, which maps \emph{type names} ($\in \mathcal{N}$\,) to
\emph{types} ($\in \mathbb{T}$). The grammar of types
(Fig.~\ref{fig:grammar-types}) is similar to C, with
the addition of {\it refinement types}~\cite{freeman1991refinement} (using predicates prefixed by
\textbf{with}).  
Type annotations apply to objects of the interface. They allow to specify
arithmetic constraints (integer values) non-nullity (pointers) and length
constraints (for arrays whose length is not known statically).

\mysubparagraph{Types as labels for regions} The types are used to
label regions in memory. For instance, in
Fig.~\ref{fig:system-memory-dump}, the region $\mathtt{a2-ab}$ has
type $\mathtt{Thread[2]}$, while $\mathtt{ae-b1}$ has type
$\mathtt{Mem\_Table}$. When an address is statically allocated, the
corresponding declaration provides the labeling (e.g.\ the addresses
$\mathtt{a2-ab}$ have been allocated by the global C declaration
\lstinline{Thread thread_array[2]}).

Formally, labels are represented using a \emph{labeling}
$\mathscr{L} \in \mathbb{A} \to \mathbb{T} \times \mathbb{N}$,
mapping addresses to (type, offset) pairs. For instance in
Fig.~\ref{fig:system-memory-dump}, the label of $\mathtt{a2}$,
$\mathscr{L}(\mathtt{a2})$, is $\mathtt{Thread[2]_0}$ (byte 0 of a
$\mathtt{Thread[2]}$ region);
$\mathscr{L}(\mathtt{a3})=\mathtt{Thread[2]_1}$ (byte 1 of a
$\mathtt{Thread[2]}$ region); etc.

\mysubparagraph{Subtyping and overlapping regions} Sometimes the regions overlap. For instance
the content at offset 3 in a $\mathtt{Thread[2]}$ is the content at offset 3 in a $\mathtt{Thread}$, which is is the content at offset
2 in a $\mathtt{Context}$, which is a $\mathtt{Flags}$. Thus, we want
the address \texttt{a6} to be labeled by all four of
$\mathtt{Thread[2]_3}$, $\mathtt{Thread_3}$, $\mathtt{Context_2}$, and $\mathtt{Flags}$.
We express this property as a \emph{subtyping} relationship:
$\mathtt{Thread_3}$ is a \emph{subtype} of $\mathtt{Context_2}$
(written $\mathtt{Thread_3} \sqsubseteq \mathtt{Context_2}$), meaning that
addresses labeled by $\mathtt{Thread_3}$ are also labeled by
$\mathtt{Context_2}$.
We found out that this subtyping relationship is very important for
programs written in machine languages: while in C we can distinguish
\lstinline{t} from \lstinline{&t->mt} (when $\mathtt{t}$ is a
$\mathtt{Thread}*$), there is no distinction between them in machine code
(because $\mathtt{mt}$ has offset 0). Thus in machine code, a pointer to
a structure must simultaneously point to its first field.

\begin{figure}

      \centering\begin{tikzpicture}[xscale=1.7,yscale=0.7]
    \node (word) at (0.5,1.2) {$\mathtt{Word}$};      
    \node (flags) at (2,-0.1) {$\mathtt{Flags}$};
    \node (memstar0) at (-1,0) {${\mathtt{Mem\_table}_{\textcolor{gray}{0}}{*}}$};
    \node (threadstar0) at (3,0) {${\mathtt{Thread}_{\textcolor{gray}{0}}{*}}$};

    \node (int) at (0.5,0.4) {$\mathtt{Int8}$};

    \node (context0) at (0,-1) {$\mathtt{Context}_0$};
    \node (context1) at (1,-1) {$\mathtt{Context}_1$};        
    \node (context2) at (2,-1) {$\mathtt{Context}_2$};

    \node (thread0) at (-1,-2) {$\mathtt{Thread}_0$};
    \node (thread1) at (0,-2) {$\mathtt{Thread}_1$};        
    \node (thread2) at (1,-2) {$\mathtt{Thread}_2$};
    \node (thread3) at (2,-2) {$\mathtt{Thread}_3$};
    \node (thread4) at (3,-2) {$\mathtt{Thread}_4$};    

    \begin{scope}[shift={(0.25,0)}]

    \node[scale=0.8] (thread20) at (-1.5,-3) {$\mathtt{Thread[2]}_0$};
    \node[scale=0.8] (thread21) at (-0.5,-3) {$\mathtt{Thread[2]}_1$};        
    \node[scale=0.8] (thread22) at (0.5,-3) {$\mathtt{Thread[2]}_2$};
    \node[scale=0.8] (thread23) at (1.5,-3) {$\mathtt{Thread[2]}_3$};
    \node[scale=0.8] (thread24) at (2.5,-3) {$\mathtt{Thread[2]}_4$};    

    \node[scale=0.8] (thread25) at (-1,-3.5) {$\mathtt{Thread[2]}_5$};
    \node[scale=0.8] (thread26) at (0,-3.5) {$\mathtt{Thread[2]}_6$};        
    \node[scale=0.8] (thread27) at (1,-3.5) {$\mathtt{Thread[2]}_7$};
    \node[scale=0.8] (thread28) at (2,-3.5) {$\mathtt{Thread[2]}_8$};
    \node[scale=0.8] (thread29) at (3,-3.5) {$\mathtt{Thread[2]}_9$};    

    \end{scope}

    \draw (int) edge (word);

    \draw (word) edge (memstar0);            
    \draw (word) edge (threadstar0);        
    \draw (flags) edge (context2);
    \draw (memstar0) edge (thread0);
    \draw (threadstar0) edge (thread4);

    \draw (int) edge (context0);
    \draw (int) edge (context1);
    \draw (int) edge (flags);

    \draw (thread1) edge (context0);
    \draw (thread2) edge (context1);
    \draw (thread3) edge (context2);        

    \draw (thread0) edge (thread20);    
    \draw (thread1) edge (thread21);
    \draw (thread2) edge (thread22);
    \draw (thread3) edge (thread23);
    \draw (thread4) edge (thread24);            

    \draw (thread0) edge (thread25);    
    \draw (thread1) edge (thread26);
    \draw (thread2) edge (thread27);
    \draw (thread3) edge (thread28);
    \draw (thread4) edge (thread29);            

  \end{tikzpicture}%
  \vspace{-5mm}
  \caption{Subtyping relations inside a $\mathtt{Thread[2]}$ region.}
  \label{fig:subtyping-relations-thread2}
  
\end{figure}
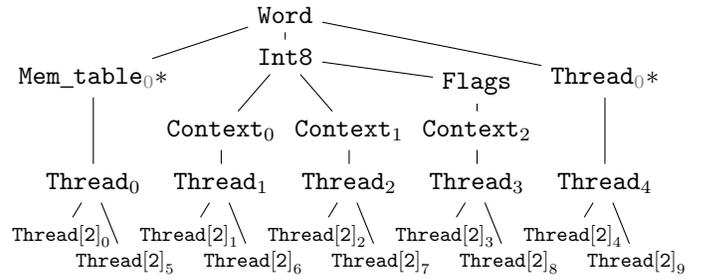

Fig.~\ref{fig:subtyping-relations-thread2} provides the subtyping
relations inside a $\mathtt{Thread[2]}$ region, derived from the
definitions of Fig.~\ref{fig:kernel-types}. To unclutter notations, we allow
ourselves to write $t$ instead of $t_{\textcolor{gray}{0}}$. We derive the
subtyping
relationships as follows: if $u$ contains a $t$ at offset $i$, then
$u_{i+k} \sqsubseteq t_k$ for all $k$ such that
$0 \le k < \operatorname{sizeof}(t)$ (with $t_0 = t$ in the case of
scalar types). In addition, for any $t$, the refined type
``$t\ \mathbf{with}\ p$'' is always a subtype of $t$.

\mysubparagraph{Separation} In our type system, two addresses whose
types are not in a subtype relationship are separated (i.e.\ do not
alias):%
\begin{multline*} \forall t_i, u_j \in \mathbb{T} \times \mathbb{N}:\ t_i \not \sqsubseteq u_j \land u_j \not \sqsubseteq t_i \Rightarrow \\
  \{ a \in \mathbb{A} : \mathscr{L}(a) \sqsubseteq t_i \} \cap \{ b \in \mathbb{A} : \mathscr{L}(b) \sqsubseteq u_j \} = \emptyset
\end{multline*}
This property ensures, for instance, that writing to an address in a
$\mathtt{Thread[2]}$ region does not modify the content of any
$\mathtt{Mem\_table}$ region.

\mysubparagraph{Types as sets of values} Types are used not only to
label addresses, but also to represent sets of values. For instance,
the set of values corresponding to $\mathtt{Int8}$ is $\mathbb{V}$,
the set of all the bitvectors of length 8; the set of values
corresponding to $\mathtt{Flags}$ is all the bitvectors with their
\textsl{PRIVILEGED} bit unset; and the set of values corresponding to
type $\mathtt{Context_0*}$ is $\{\mathtt{a3},\mathtt{a8}\}$, i.e.\ the
set of addresses that have $\mathtt{Context_0}$ as label. Note that
the definition of this latter set relies on the labeling
$\mathscr{L}$.

Formally, we provide an \emph{interpretation} $\llparenthesis \cdot \rrparenthesis_{\mathscr{L}} \in
\mathbb{T}\times\mathbb{N} \to \mathcal{P}(\mathbb{V})$, mapping a (type,
offset) pair to a set of values as follows.
\vspace{-2mm}
\begin{align*}
  \llparenthesis \mathtt{Int8} \rrparenthesis_{\mathscr{L}} =   \llparenthesis \mathtt{Word} \rrparenthesis_{\mathscr{L}}  &= \mathbb{V} \\
  \llparenthesis {t_i*} \rrparenthesis_{\mathscr{L}} &= \{ a \in \mathbb{A} : \mathscr{L}(a) \sqsubseteq t_i \} \\
  \llparenthesis {t \mathbf{\ with\ } p} \rrparenthesis_{\mathscr{L}} &= \{ x \in   \llparenthesis t \rrparenthesis_{\mathscr{L}} : p(x) \} \\  
  \llparenthesis s_k \rrparenthesis_{\mathscr{L}} &= \bigcap_{\substack{t\in\mathbb{T} \\[1pt] s_k \sqsubseteq t_i}} \llparenthesis t_i \rrparenthesis_{\mathscr{L}}\text{ for other types}
\end{align*}

\mysubparagraph{Well-typedness} The most important property of this
type system is, given a set of types like in
Fig.~\ref{fig:kernel-types}, the \emph{well-typedness} of a memory $m$ with
regard to a labeling $\mathscr{L}$, defined as:
\[ \forall a \in \mathbb{A},\quad m[a] \in \llparenthesis \mathscr{L}(a)
  \rrparenthesis_{\mathscr{L}} \] which means that the contents of
memory $m$ at address $a$ should be one of the admissible values
for the type label of $a$. For instance in
Fig.~\ref{fig:system-memory-dump}, the address $\mathtt{a6}$ is
labelled as $\mathtt{Thread[2]}_4$, which is a subtype of
$\mathtt{Thread}_0*$; the only addresses tagged with a subtype of
$\mathtt{Thread}_0$ are $\mathtt{a2}$ and $\mathtt{a7}$, so the memory
at $\mathtt{a6}$ can contain only $\mathtt{a2}$ or $\mathtt{a7}$; this
is correct in Fig.~\ref{fig:system-memory-dump}. Conversely, if
$m[\mathtt{a6}]$ were equal to $\mathtt{ae}$, the couple
$(m,\mathscr{L})$ would not be well-typed.

\mysubparagraph{The type-based shape abstract domain} Our type-based
shape abstract domain does not track the contents of the rest of user
image memory at all (which makes it very efficient). Instead, it
guarantees that each operation \emph{preserves well-typedness}:
i.e.\ that modifying a memory $m$ for which $\mathscr{L}$ is a
labeling, results in a new memory $m'$, for which $\mathscr{L}$ is
still a labeling. This is a useful property: it means that after a
store, a register with type $\mathtt{Task_0*}$ will still point to a
$\mathtt{Task}$ structure, and not, e.g., to a page table. To do so, it
tracks the types that are contained in the registers and kernel memory
(i.e.\ stack, global variables, etc.) at each program point.

Let us return to the invariant computed by the parameterized analysis
(Section~\ref{sec:illustr-example-kern}). The exact meaning of the
invariant is that $(m,\mathscr{L})$ exists such that $(m,\mathscr{L})$
is well-typed; that the memory at address $\mathtt{a0}$ contains
values in
$\llparenthesis \mathtt{Thread *} \rrparenthesis_{\mathscr{L}}$; the
memory at address $\mathtt{a1}$ contains values in
$\llparenthesis \mathtt{Context *} \rrparenthesis_{\mathscr{L}}$; etc.
The precise user memory and labeling at
Fig.~\ref{fig:system-memory-dump} is indeed a special case of this
invariant.

\section{Case study \& Experimental Evaluation}
\label{sec:case-study}

We seek to answer the following Research Questions:

\begin{itemize}[leftmargin=2.5em,nosep]
\item[\textbf{RQ0:}] \textbf{Soundness check} Does our analyzer fail
  to verify APE and ARTE when the kernel is vulnerable or buggy?
\item[\textbf{RQ1:}] \textbf{Real-life Effectiveness} Can our method 
  verify real (unmodified) embedded kernels? %
\item[\textbf{RQ2:}] \textbf{Internal evaluation} What are the respective impacts of the
  different elements of our method?    
\item[\textbf{RQ3:}] \textbf{Genericity} Can our method apply on different kernels, hardware architectures and toolchains?
\item[\textbf{RQ4:}] \textbf{Automation} Is it possible to prove APE and ARTE in
  OS kernels fully automatically?
\item[\textbf{RQ5:}] \textbf{Scalability} Can our method scale to large %
numbers of tasks?
\end{itemize}

\subsection{Experimental setup}

\mysubparagraph{\osname{}} We consider for RQ1 and RQ2 the 
\osname{} kernel, an {\it industrial}  solution
for implementing security- and safety-critical hard real-time
applications, used in industrial automation, automotive, aerospace and
defense. The kernel is  developed by
\company\Ldash{}an SME whose engineers are not formal method
experts\Rdash{}using standard compiler toolchains.

We consider a port of the kernel to a \emph{4-cores} ARM
Cortex-A9 processor with \emph{ARMv7} instruction set. It relies on  ARM MMU for memory
protection (\emph{pagination}). The kernel features a hard real-time scheduler that dispatches the tasks between the cores (migrations are allowed), and monitors timing budgets and deadlines. The kernel adopts a ``static microkernel'' architecture, with unprivileged services used to monitor inter-process communication. 
The configuration for the user tasks is generated by the \osname{} toolchain, and all the memory is statically allocated. The system is
parameterized: the kernel and the user images are compiled separately
and both are loaded at runtime by the bootloader. %
We have analyzed two versions: 
\begin{itemize}[nosep]%
\item \textbf{\textsc{beta}}, \textls[-14]{a preliminary version where  
we found a \mbox{vulnerability};} %

\item \textbf{\textsc{v1}}, a more polished version with the
  vulnerability fixed and debug code was removed.
\end{itemize}

The code segment of each kernel executable contains 329 functions
($\mathtt{objdump}$ reports around 
$10,\!000$ %
instructions), shared between the kernel and the core unprivileged services.
Finally, \company{} provided us with a sample user image. 

\mysubparagraph{EducRTOS} For research questions where we need 
some flexibility 
(RQ0, RQ3--5), we developed a new embedded
kernel called EducRTOS. It contains a
variety of features (e.g.\ different schedulers, dynamic thread
creation) complementary to those of \osname{} (e.g.\ x86 instead of ARM,
segmentation instead of pagination). 
EducRTOS is also being used to teach operating systems
to master students. Depending on the included features, the kernel size ranges
between 2,346 and 2,866 instructions.

\mysubparagraph{Implementation} Our static analysis method is implemented in a prototype named \binseccodex{}:
a plugin (41k lines of OCaml) on top of the open source
\binsec{}~\cite{david_binsec/se:_2016} framework for binary-level semantic
analysis. We reuse the ELF parsing, instruction decoding, and intermediate
representation lifting of the platform \cite{bardin2011bincoa} and have
reimplemented a whole static analysis on top of it. Since the analysis is
performed on \binsec's intermediate representation, the analysis implementation is entirely
independent from the hardware architecture, apart from the empowered transition.
To implement this transition, we use a simple sound approximation consisting in setting to an arbitrary value any unprotected
register or memory location. Its implementation takes 23~lines of OCaml for ARM
and 62~lines for x86.

\mysubparagraph{Availability} \binseccodex{} and \educrtos{}
are open-source and part of the artifact accompanying this paper, but
\osname{} is a commercial product.

\mysubparagraph{Experimental conditions} %
We performed our formal verification
completely independently from \company{} activities. In particular, we
never saw the source code of their kernel, and our interactions with
\company{} engineers were limited to a general presentation of \osname{}
features. %
We ran all our analyses on a
standard laptop with an Intel Xeon E3-1505M 3~GHz CPU with 32 GB
RAM. All measurements of execution time or memory usage %
have been 
observed to vary by no more than 2\,\% across 10 runs. We took the mean value of
the runs.

\subsection{Soundness check (RQ0)}

Our method is sound {\it in principle}, in that it proves by construction APE and ARTE
only on kernels where these properties are true. Yet, 
this ultimately depends on the correct implementation of the static analysis; in this
preliminary experiment, we want to check that our tool indeed does not prove
APE and ARTE on buggy kernels.

\mysubparagraph{Protocol} We deliberately introduce 4
backdoors in \educrtos{} by creating 4 new system calls that
jump to an arbitrary code address with kernel privilege, grant kernel privilege
to user code segments, write to an arbitrary address, or modify the memory
protection tables to cover parts of the kernel address space.  These backdoors
can easily be exploited to gain control over the kernel. Additionally, we add 
3 bugs possibly leading to crashes in existing system calls: a read at an
arbitrary address, an illegal opcode error and a possible division by zero.

\mysubparagraph{Results} For each of the added vulnerabilities, our
analysis does not prove APE (it either failed or reported an alarm at
the instruction where the error occurs); for each bug, it does not
prove ARTE.%

\mysubparagraph{Conclusions} \binseccodex{} was able to detect all the
privilege escalation vulnerabilities and runtime errors that we
explicitly introduced.

\mysubparagraph{Additional notes} This corresponds to the experience
we had while developing EducRTOS, where several times we launched
\binseccodex{} and discovered unintentional bugs  (wrong check on
the number of syscalls, write to null pointers) that were not detected
by testing. Our method even discovered a bitflip in a kernel
executable that occurred when the file was copied. %

\subsection{Real-Life Effectiveness (RQ1)}

\mysubparagraph{Protocol} %
Our goal here is to
evaluate the \emph{effectiveness} of our parameterized verification
method on an unmodified industrial kernel, measured by:
(1) the
fact that the {\it method indeed  succeeds} in computing a non-trivial invariant for the
whole system, i.e., computes an  {invariant under  precondition} for the kernel runtime and checks
that the user tasks establish the precondition;  
(2) the \emph{precision}
of the analysis, measured by the number of \emph{alarms}
(i.e.\ properties that the analyzer cannot prove); 
(3) the \emph{effort} 
necessary to setup the analysis, measured by the number of lines of
manual annotations; %
 and 
(4) the \emph{performance} of the analysis,
measured in CPU time and memory utilization. 

The bulk of the annotations (1057 lines) consists in type definitions, and was
automatically extracted from the sample user image, using debug information in the user executable.
Debug information is otherwise not used by the analysis (and the kernel
executable does not contain any). Manual annotations consist in changing the definitions by adding ``\textbf{with}''
predicates, or information about the length of arrays. These annotations were reverse-engineered as the ones required for the kernel analysis to succeed with no remaining alarm (and also
correspond to necessary requirements for the code to run without errors),
but they could be obtained from the documentation or directly provided by the OS
developers.

We consider both \osname{} kernel versions and two configurations (i.e., sets of
type annotations): 
\begin{itemize}
\item  \textbf{\textit{Generic}} contains types and parameter invariants which must hold for all
legitimate user images;  %

\item \textbf{\textit{Specific}} further assumes that the stacks of all user
  tasks in the image have the same size. 
  This is the default for \osname{} applications, and it holds on our case
  study. 
   
\end{itemize}

\begin{table}[tbp]
  \centering
  \caption{Main verification results on \osname}
  \vspace{1mm}
  \label{tab:main-verif-results}
  \newcolumntype{"}{@{\hskip\tabcolsep\vrule width 0.8pt\hskip\tabcolsep}}  
  \begin{tabular}{|p{14.5mm}|p{10mm}|c|c"c|c|}
    \cline{3-6}
    \multicolumn{2}{c|}{} & \multicolumn{2}{c"}{\emph{Generic}} & \multicolumn{2}{p{20mm}|}{\hspace{-1mm}\makebox[20mm]{\emph{Specific}\phantom{\rule{0pt}{1em}}}} \\
    \hline
     \multirow{2}{*}{\hspace{0mm}\makecell[r]{\# shape\\annotations}} &\makebox[10mm]{converted}& \multicolumn{4}{c|}{1057\phantom{\rule{0pt}{1em}}} \\
    \cline{2-6}
    & \phantom{\rule{0pt}{1em}}{manual} & \multicolumn{2}{c"}{57 (5.11\%)} & \multicolumn{2}{c|}{58 (5.20\%)}\\    
    \hline\hline
    \multicolumn{2}{|c|}{\makecell[c]{Kernel version}\phantom{\rule{0pt}{1em}}} & \makecell{\textsc{beta}} & \makecell{\textsc{v1}} & \makecell{\textsc{beta}} & {\makebox[1mm]{\textsc{v1}}}\\
    \hline
    \multirow{2}{*}{\hspace{-1.3mm}\makecell[r]{invariant\\computation}} & \makecell[r]{status} & \phantom{\rule{0pt}{1em}} \tick & \tick & \tick & \tick \\
    \cline{2-6}
    & \makebox[8mm]{\quad time (s)} & \phantom{\rule{0pt}{1em}} 647 & 417 & 599 & \makebox[0mm]{406} \\
    \hline
    \multicolumn{2}{|c|}{\makecell[c]{\# alarms in runtime}} & \makecell{\makebox[11mm]{\scriptsize 1 \textbf{true error}}\\\makebox[6mm]{\scriptsize 2 false alarms}} & \makecell{\makebox[4mm]{\scriptsize 1 false}\\\scriptsize\makebox[4mm]{alarm}} & \makecell{\makebox[10mm]{\scriptsize 1 \textbf{true error}}\\\makebox[10mm]{\scriptsize 1 false alarm}} & \makebox[0mm]{\textbf{\scriptsize 0} \tick}  \\
    \hline
    \multirow{2}{*}{\hspace{1.5mm}\makecell[r]{user tasks\\checking}} & {\makecell[r]{status}} & \phantom{\rule{0pt}{1em}} \tick & \tick & \tick & \tick \\ \cline{2-6}

    & \makebox[8mm]{\quad time (s)} & \phantom{\rule{0pt}{1em}} 32 & 29 & 31 & \makebox[0mm]{30} \\[1pt]
    \hline
    \hline

    \multicolumn{2}{|c|}{Proves APE and ARTE?} & \phantom{\rule{0pt}{1em}} N/A  & \textcolor{orange!50!green!50!black!50}{\large $\mathbf{\sim}$}  & N/A  & \tick  \\[1pt]
    \hline
  \end{tabular}
\end{table}

\mysubparagraph{Results} The main results are given in
Table~\ref{tab:main-verif-results}. The \textbf{\textit{generic}} annotations consist in only 
57 lines of manual annotations, in addition to 1057 lines that were automatically
generated (i.e.\@ 5\% of manual annotations, and a manual annotations per instructions ratio of 0.58\%). The \textbf{\textit{specific}} annotations adds one more line. When analyzing the \textbf{\textsc{beta} version} with these annotations,
only {\it 3 alarms} are raised in the runtime:

\begin{itemize}
\item One is a {\bf true vulnerability}: in the supervisor call entry
  routine (written in manual assembly), the kernel extracts the system
  call number from the opcode that triggered the call, sanitizes it
  (ignoring numbers larger than 7), and uses it as
  an index in a table to jump to the target system call function; but
  this table has only 6 elements, and is followed by a string in
  memory. This off-by-one error allows an $\mathtt{svc\ 6}$ system call to
  jump to an unplanned (constant) location, which can be
  attacker-controlled in some user images. The error is detected as
  the target of the jump goes to a memory address whose content is not
  known precisely, and that we thus cannot~decode;

\item One is a false alarm caused by {\it debugging code} temporarily
  violating the shape constraints: the code writes the constant
  $\mathtt{0xdeadbeef}$ in a memory location that should hold a pointer to a user stack
  (yielding an alarm as we cannot prove that this
  constant is a valid address for this type), and that
  memory location is always overwritten with a correct value further
  in the execution;

\item  The last one is a false alarm caused by an imprecision in our 
  analyzer when user stacks can have different sizes.
\end{itemize}

When analyzing the  \textbf{v1 version}, the first two alarms disappear (no new
alarm is added). Analyzing the kernel with the \textbf{\textit{specific}} annotations   %
makes the last alarm disappear. In all cases user tasks checking succeeds. 
\begin{sectionhighlight}
\textls[-12]{\textit{Analyzing the \textbf{v1} kernel with the \textbf{\textit{specific}}
annotations allows to reach 0 alarms, meaning that we have formally verified APE and ARTE.}}
\end{sectionhighlight}

Computation time is {\it low}: less than 11 minutes for
the parameterized analysis, and 35 seconds for base case
checking. %

\mysubparagraph{Conclusions} This experiment shows that \emph{it is feasible to verify
  absence of privilege escalation of an industrial microkernel using
  only fully-automated methods} (albeit with a very slight
amount of manual annotations), and \emph{without any change to the original kernel}. Especially:

\begin{itemize}
\item The analysis is {\it effective}, in that it identifies real errors in the code, and verifies
  their absence once  removed; 

\item The analysis is {\it very precise}, as we manage  to reach  0 false alarms 
on the correct code, and we had no more than 2 false alarms on each configuration of the analysis; 

\item The \emph{annotations burden is very small} (58 simple lines),
  as the kernel invariant is computed automatically and most of the
  shape annotations are automatically converted from the interface types;

\item {Finally, the \emph{analysis time}, for a kernel whose size is
  typical for embedded microkernels, \emph{is small} (between 406 and 647 seconds).
  Analysis of CPU$_0$ alone takes only 86 seconds.}
\end{itemize}

\subsection{Evaluation of the method (RQ2, sketch only) }

We performed a detailed evaluation of the method, 
described in full in the supplementary data\footnote{\url{https://binsec.github.io/assets/publications/papers/2021-rtas-supplementary-data.pdf}}.
This experiment consists in evaluating, on
the \textbf{\textsc{v1}} kernel version, our 3-step method
(Section~\ref{sec:3-step-method}), using different configurations for
the weak shape domain.
Results show that the
\emph{shape domain is necessary for a parameterized verification of
  the kernel} (otherwise the analysis aborts due to imprecision).  While 1057 lines of type annotations are automatically
converted from the types for the user image interface, only \emph{10
  manual additional lines are necessary} for the analysis to
\emph{succeed in computing an invariant} (and 58 to eliminate
all the alarms in the runtime). %

\subsection{Genericity (RQ3)}

\mysubparagraph{Protocol} 
We applied
our method  both to \osname{} and \educrtos{}, two kernels running on different
architectures (resp. ARM and x86) and memory protection mechanisms (resp.\ segmentation and pagination).
We also performed  a parameterized verification of 96 \educrtos{}
variants. Each variant was compiled with a different valuation of the
following parameters:
\begin{itemize}
  \item compiler: GCC 9.2.0, Clang 7.1.0.
  \item optimization flags: \texttt{-O1}, \texttt{-O2}, \texttt{-O3} or
    \texttt{-Os}.
  \item scheduling algorithms: round-robin, fixed-priority, or
    earliest-deadline-first scheduling.
  \item dynamic thread creation: enabled or disabled.
  \item debug printing: enabled or disabled.
\end{itemize}

\mysubparagraph{Results} We could {\it parametrically} verify all  96 \educrtos{} variants.
The verification requires 98 lines of type definitions (which we extracted
automatically, as we did for \osname, from debug information) and 12 to 14 lines of manual annotations (depending on the
scheduler), with 12 lines being common to all variants; corresponding to an annotation per instruction ratio of less than 0.59\%. %
The verification of each variant takes between 1.6\,s and 73\,s. The detailed measurements for each variant are provided in the
supplementary data.

\mysubparagraph{Conclusions} Our method is applicable to different
kernel features and hardware architectures. It is robust: non-trivial
changes in the kernel executable, like introduction of new features,
or change in code patterns because of compiler or changes in
optimisation levels, do not not require to change the
configuration. This robustness makes it an interesting tool for
verifying kernels coming  in many variants.

\subsection{Automation (RQ4) and Scalability (RQ5)}

\mysubparagraph{Protocol} In this experiment, we perform a
fully automated in-context verification (with {\it no annotation}) of a simple variant of the \educrtos{}
kernel with round-robin scheduling and no dynamic task creation.
We then use this kernel to evaluate the scalability of
this approach by modifying the number of tasks that run on the
system, and compare it against our parameterized method. %

\begin{figure}[htbp]
  \hspace{-4mm}\makebox{\input{data.tikz}}
\vspace{-8mm}  
\caption{Performance when verifying a system with $N$ tasks}%
\label{fig:scalability}
\end{figure}
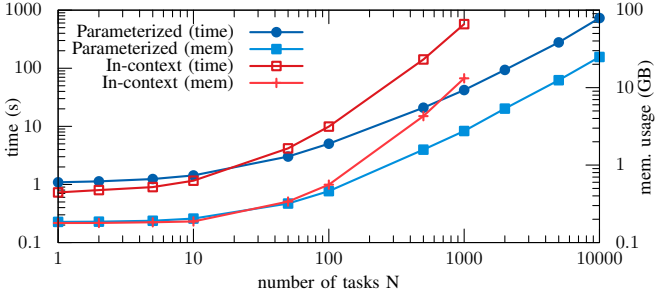

Figure~\ref{fig:scalability} provides the
execution time and memory used when verifying a system composed of $N$
tasks for both approaches. For the parameterized analysis, the
invariant computation takes less than a second, and total analysis time
is almost linear in the number of tasks, while requiring only 12 lines of annotations. In this case,
the only parts that depends on the number of tasks are the base case checking and the
in-context static analysis of the almost deterministic boot code.

By contrast, for the in-context verification, the number of steps to
reach a fixpoint, the number of modified memory locations, and the
number of target locations for the $\mathtt{Thread *}$ pointers grow
with the number of tasks, resulting in a quadratic complexity. This
scalability issue is inherent to the use of in-context
verifications~\cite{nordholz2020design}.

\mysubparagraph{Conclusions} Fully-automated in-context verification
with no annotation is achievable on very simple kernels, but is not
robust enough for more complex kernels. Moreover, it does not scale
to very large number of tasks. On the contrary, parameterized
verification (with very few annotations) is robust and scales almost linearly with large numbers of tasks.

\section{Related Work}

There is a long history of using formal methods to verify operating
system kernels.

\mysubparagraph{Degree of automation} %
We can distinguish three classes of verification methods: 

\medskip
\begin{itemize}[nosep]
\item \emph{manual}
  \cite{bevier1989kit,richards2010modeling,gu2015deep,xu2016practical,klein2009sel4}: the user has to 
  {\it provide for every program point a candidate invariant}, 
  then {\it prove} via a
  proof assistant that every instruction preserves these candidate invariants;

\item \emph{semi-automated}\footnote{Some authors call these techniques automated; we use the word semi-automated to emphasize the difference with fully-automated methods.} 
  \cite{alkassar2010automated,yang2010safe,dam2013machine,vasudevan2013design,vasudevan2016uberspark,ferraiuolo2017komodo,nelson2019scaling,nordholz2020design}: the user has to 
  {\it provide} the candidate invariants {\it at some key program 
  points} (kernel entry and exit, loops, and optionally other program points like function entry and exit) and then 
  use automated provers to verify that all %
  paths between these
  points preserve the candidate invariants; 

\item \emph{fully automated}: a sound static analyzer~\cite{cousot1977abstract} %
  {\it automatically infers} correct invariants for every program
  point. The user only {\it provides invariant templates} by selecting
  or configuring the required abstract domains. 
  This
  approach has been used to verify source code of critical embedded
  software \cite{blanchet2003static}, but never to 
  verify a kernel.
\end{itemize}

\begin{sectionhighlight}
  We demonstrate that abstract interpretation can
  automatically verify embedded kernels from their executable with a very low annotation burden, 
  sometimes with no manual intervention.
\end{sectionhighlight} 

In the parameterized case, we verify kernels with a ratio of
annotations per instruction which is at worst of 13 lines for 2346
instructions (0.59\%). By comparison, the automated verification of
CertiKOS$^S$~\cite{nelson2019scaling} required 438 lines for 1845 instructions (23.7\%).

\mysubparagraph{Target property} Prior kernel verification methods 
generally target  four kinds of program properties:
\medskip
\begin{itemize}[nosep]
\item \emph{Functional correctness}\cite{walker1980specification,bevier1989kit,klein2009sel4,alkassar2010automated,paul2012completing,klein2014comprehensive,gu2015deep,xu2016practical,ferraiuolo2017komodo} i.e.\ compliance
  to a (manually written) formal specification of the kernel; 

\item \emph{Task separation}, \cite{rushby1981design,nelson2019scaling}
  i.e.\ verifying the absence of undesirable %
  information flow between  tasks; 

\item \emph{Absence of privilege escalation}
  \cite{yang2010safe,vasudevan2016uberspark,nordholz2020design} (a.k.a.~ 
   kernel integrity), i.e.\ proving that the kernel protects
  itself and that no attacker can gain control over it;
\item \emph{Absence of runtime
    errors}~\cite{vasudevan2013design} like
  buffer overflows, null-pointer dereferences, or format string
  vulnerabilities.
\end{itemize}%
\begin{sectionhighlight}
  To achieve maximal automation, 
  we focus on \emph{implicit}
  properties. %
  Especially, we are the first to prove that  absence of privilege escalation is implicit. 
\end{sectionhighlight}

Note  that the invariants we infer 
could significantly reduce the number of annotations
required to verify functional correctness~\cite{nelson2019scaling}, and can generally help any other analysis; for instance
worst-case execution time (WCET) estimation~\cite{schuster2019proving,colin2001wcet}
requires knowledge about the CFG and memory accesses. Stronger invariants can be inferred by
combining our analysis with other domains~\cite{cousot1979systematic}.

\mysubparagraph{Trusted computing base (TCB) and verification comprehensiveness} %
We only trust that the bootloader correctly loads the ELF image in
memory, that the hardware complies with its specification, and that
our abstract interpreter (which has no dependencies) is sound. We do
not trust the build toolchain (compiler, build scripts, assembler and
linker), we analyze all of the code, and we do not assume any
unverified hypothesis.

While push-button methods can verify all of the
code~\cite{dam2013machine,nelson2019scaling,nordholz2020design}, more
manual methods
often~\cite{xu2016practical,walker1980specification,klein2014comprehensive,gu2016certikos}
leave  parts of the code unverified when the verification is hard or
overly tedious. %
While source-level verifications methods sometimes carry to the assembly
level~\cite{sewell2013translation} or include support for assembly
instructions~\cite{paul2012completing,vasudevan2016uberspark}, they 
usually trust the compilation, assembly and linking phases. 

To further reduce our TCB, we would have to verify our static
analyzer in a proof language with a small
kernel (like Isabelle \cite{nipkow2002isabelle} or Coq \cite{coquand1988calculus}), a huge effort that has been shown to be feasible~\cite{jourdan2015formally}.   %

\mysubparagraph{Features of verified kernels} 
 We focus on embedded systems kernels, where the
 kernel memory is mostly %
 statically allocated (either in the kernel or in the
 user image)
 (e.g.\ \cite{ramamritham1994scheduling,muehlberg2011verifying,ARINC653,richards2010modeling,dam2013machine,nelson2019scaling,nordholz2020design}). 
We do not consider real-time executives where memory protection is absent or optional, as verifying APE on them would require unchecked assumptions on applicative code. 
 Still, the kernels that we have analyzed feature complex code,
 including dynamic thread creation and dynamic memory allocation using
 object pools; different memory protection tables (using x86 segments
 or ARM page tables) modified at boot and runtime; different real-time
 schedulers working on arbitrary numbers of tasks; boot-time
 parametrization by a user image; and usage of multiple cores with
 shared memory. %

\begin{sectionhighlight}
  Being based on abstract interpretation, our method  can address kernels
  with features out of reach of prior attempts based on symbolic execution 
  (e.g., real-time schedulers). While we can still design kernels to
  get around the limitations of the tool~\cite{nelson2019scaling}, 
  {\it we are the first to verify an existing, unmodified real kernel}. 
\end{sectionhighlight}

However, like any sound static analyzer, \binseccodex{} may still be too
imprecise on some code patterns, emitting {\it false alarms} or
failing to compute a non-trivial invariant. For instance, our tool
would not handle well self-modification in the kernel, complex
lock-free code, or using a general-purpose memory allocator outside of
the boot code. Still, we can directly reuse progress in the automated
analysis of these patterns. %
An important pattern for real-time systems is full preemptibility in the kernel.
It should be possible to extend our analysis to these systems by leveraging
our capacity to handle concurrent executions, notably by adding
a stack abstraction that would be independent
from the concrete address (as in \cite{reps2010there}); this is an important topic
for future work.

\mysubparagraph{Verifying systems with unbounded memory} Prior works
targeting  machine code
\cite{bevier1989kit,dam2013machine,nelson2019scaling,nordholz2020design}
use a flat representation of the memory (where the abstract state
enumerates all the memory cells in the kernel), causing 
scalability issues in the presence of a large number of
tasks~\cite{nordholz2020design}  and preventing \emph{parameterized}
verification. 
To handle systems where the amount of memory is not statically known,
we need more complex representations that  \emph{summarize} memory.
Points-to and alias analyses are fast and easy to setup but are too imprecise
for formal verification, and generally assume that the code behaves
nicely, e.g., type-based alias analyses~\cite{diwan1998tbaa} assume
that programs comply with the strict aliasing rule -- while kernel codes 
often do not conform to C standard \cite{klein2014comprehensive}. 
On the other hand,  shape
analyses~\cite{sagiv1999parametric,chang2008relational} can fully
prove memory invariants, but require heavy configuration and
 generally cannot  scale to a full embedded kernel. 

\begin{sectionhighlight}
We propose a weak type-based shape abstract domain that hits a middle ground: it is fast, precise,  
handles low-level behaviors (outside of the C standard) and requires little
configuration. This is also the first time a shape analysis is performed on machine code.
\end{sectionhighlight}

{Marron~\cite{marron2012structural} also describes a
weak shape domain, but on a type-safe
language with no implicit type casts, pointer arithmetic, nor nested
data structures.}  
Outside of fully-automated analyses, Walker et al.~\cite{walker1980specification} already observed in the 1980s 
that reasoning on type invariants is well suited to OS kernel
verification. Several systems build around this idea \cite{yang2010safe,ferraiuolo2017komodo}, leveraging a dedicated typed language.
Cohen et al.~\cite{cohen2009precise} describe a typed semantics for C
with additional checks for memory typing preservation, similar to our
own checks on memory accesses.  While they use it in a deductive
verification tool for C (to verify a hypervisor
\cite{alkassar2010automated}), we build an abstract interpreter for
machine code.

\section{Conclusion and future work}

We have presented \binseccodex{}, an automated method for formally
verifying embedded kernels (absence of runtime errors and absence of
privilege escalation) directly from their executable, with only a very low annotation burden.  
We address
important limitations of existing automated methods: we allow
\emph{parameterized} verification, i.e.\ verifying a kernel
independently from the applications running on it; we handle unbounded
loops that are necessary for implementing real-time schedulers; and we
\emph{infer} the kernel invariants, instead of merely checking
them. As in OS formal verification, {\it ``invariant reasoning
  dominates the proof effort''}\cite{walker1980specification} (in
seL4, 80\% of the effort was spent stating and verifying
invariants~\cite{klein2009sel4}), this work is a key enabler for more
automated verification of more complex systems.

\section*{Acknowledgements}

The authors warmly thank Gilles Muller, Amit Vasudevan, Pierre-Yves Piriou and
Guerric Chupin, as well as the anonymous reviewers, for their very helpful remarks and discussions. Xavier Rival received funding from the French ANR,
as part of the Veriamos grant. Matthieu Lemerre and Sébastien Bardin also
received funding from the ANR as part of the TAVA grant.

\bibliographystyle{ieeetr}
\bibliography{biblio}

\end{document}

%% file: data.tikz
\begin{tikzpicture}[gnuplot]
%% generated with GNUPLOT 5.2p8 (Lua 5.3; terminal rev. Nov 2018, script rev. 108)
%% Wed May 27 21:34:49 2020
\tikzset{every node/.append style={scale=0.70}}
\path (0.000,0.000) rectangle (9.500,4.000);
\gpcolor{color=gp lt color border}
\gpsetlinetype{gp lt border}
\gpsetdashtype{gp dt solid}
\gpsetlinewidth{1.00}
\draw[gp path] (1.054,0.691)--(1.234,0.691);
\node[gp node right] at (0.925,0.691) {$0.1$};
\draw[gp path] (1.054,0.924)--(1.144,0.924);
\draw[gp path] (1.054,1.060)--(1.144,1.060);
\draw[gp path] (1.054,1.156)--(1.144,1.156);
\draw[gp path] (1.054,1.231)--(1.144,1.231);
\draw[gp path] (1.054,1.293)--(1.144,1.293);
\draw[gp path] (1.054,1.344)--(1.144,1.344);
\draw[gp path] (1.054,1.389)--(1.144,1.389);
\draw[gp path] (1.054,1.429)--(1.144,1.429);
\draw[gp path] (1.054,1.464)--(1.234,1.464);
\node[gp node right] at (0.925,1.464) {$1$};
\draw[gp path] (1.054,1.697)--(1.144,1.697);
\draw[gp path] (1.054,1.833)--(1.144,1.833);
\draw[gp path] (1.054,1.929)--(1.144,1.929);
\draw[gp path] (1.054,2.004)--(1.144,2.004);
\draw[gp path] (1.054,2.066)--(1.144,2.066);
\draw[gp path] (1.054,2.117)--(1.144,2.117);
\draw[gp path] (1.054,2.162)--(1.144,2.162);
\draw[gp path] (1.054,2.202)--(1.144,2.202);
\draw[gp path] (1.054,2.237)--(1.234,2.237);
\node[gp node right] at (0.925,2.237) {$10$};
\draw[gp path] (1.054,2.470)--(1.144,2.470);
\draw[gp path] (1.054,2.606)--(1.144,2.606);
\draw[gp path] (1.054,2.702)--(1.144,2.702);
\draw[gp path] (1.054,2.777)--(1.144,2.777);
\draw[gp path] (1.054,2.839)--(1.144,2.839);
\draw[gp path] (1.054,2.890)--(1.144,2.890);
\draw[gp path] (1.054,2.935)--(1.144,2.935);
\draw[gp path] (1.054,2.975)--(1.144,2.975);
\draw[gp path] (1.054,3.010)--(1.234,3.010);
\node[gp node right] at (0.925,3.010) {$100$};
\draw[gp path] (1.054,3.243)--(1.144,3.243);
\draw[gp path] (1.054,3.379)--(1.144,3.379);
\draw[gp path] (1.054,3.475)--(1.144,3.475);
\draw[gp path] (1.054,3.550)--(1.144,3.550);
\draw[gp path] (1.054,3.612)--(1.144,3.612);
\draw[gp path] (1.054,3.663)--(1.144,3.663);
\draw[gp path] (1.054,3.708)--(1.144,3.708);
\draw[gp path] (1.054,3.748)--(1.144,3.748);
\draw[gp path] (1.054,3.783)--(1.234,3.783);
\node[gp node right] at (0.925,3.783) {$1000$};
\draw[gp path] (1.054,0.691)--(1.054,0.871);
\draw[gp path] (1.054,3.783)--(1.054,3.603);
\node[gp node center] at (1.054,0.475) {$1$};
\draw[gp path] (1.596,0.691)--(1.596,0.781);
\draw[gp path] (1.596,3.783)--(1.596,3.693);
\draw[gp path] (1.912,0.691)--(1.912,0.781);
\draw[gp path] (1.912,3.783)--(1.912,3.693);
\draw[gp path] (2.137,0.691)--(2.137,0.781);
\draw[gp path] (2.137,3.783)--(2.137,3.693);
\draw[gp path] (2.312,0.691)--(2.312,0.781);
\draw[gp path] (2.312,3.783)--(2.312,3.693);
\draw[gp path] (2.454,0.691)--(2.454,0.781);
\draw[gp path] (2.454,3.783)--(2.454,3.693);
\draw[gp path] (2.575,0.691)--(2.575,0.781);
\draw[gp path] (2.575,3.783)--(2.575,3.693);
\draw[gp path] (2.679,0.691)--(2.679,0.781);
\draw[gp path] (2.679,3.783)--(2.679,3.693);
\draw[gp path] (2.771,0.691)--(2.771,0.781);
\draw[gp path] (2.771,3.783)--(2.771,3.693);
\draw[gp path] (2.853,0.691)--(2.853,0.871);
\draw[gp path] (2.853,3.783)--(2.853,3.603);
\node[gp node center] at (2.853,0.475) {$10$};
\draw[gp path] (3.395,0.691)--(3.395,0.781);
\draw[gp path] (3.395,3.783)--(3.395,3.693);
\draw[gp path] (3.712,0.691)--(3.712,0.781);
\draw[gp path] (3.712,3.783)--(3.712,3.693);
\draw[gp path] (3.937,0.691)--(3.937,0.781);
\draw[gp path] (3.937,3.783)--(3.937,3.693);
\draw[gp path] (4.111,0.691)--(4.111,0.781);
\draw[gp path] (4.111,3.783)--(4.111,3.693);
\draw[gp path] (4.253,0.691)--(4.253,0.781);
\draw[gp path] (4.253,3.783)--(4.253,3.693);
\draw[gp path] (4.374,0.691)--(4.374,0.781);
\draw[gp path] (4.374,3.783)--(4.374,3.693);
\draw[gp path] (4.478,0.691)--(4.478,0.781);
\draw[gp path] (4.478,3.783)--(4.478,3.693);
\draw[gp path] (4.570,0.691)--(4.570,0.781);
\draw[gp path] (4.570,3.783)--(4.570,3.693);
\draw[gp path] (4.653,0.691)--(4.653,0.871);
\draw[gp path] (4.653,3.783)--(4.653,3.603);
\node[gp node center] at (4.653,0.475) {$100$};
\draw[gp path] (5.194,0.691)--(5.194,0.781);
\draw[gp path] (5.194,3.783)--(5.194,3.693);
\draw[gp path] (5.511,0.691)--(5.511,0.781);
\draw[gp path] (5.511,3.783)--(5.511,3.693);
\draw[gp path] (5.736,0.691)--(5.736,0.781);
\draw[gp path] (5.736,3.783)--(5.736,3.693);
\draw[gp path] (5.910,0.691)--(5.910,0.781);
\draw[gp path] (5.910,3.783)--(5.910,3.693);
\draw[gp path] (6.053,0.691)--(6.053,0.781);
\draw[gp path] (6.053,3.783)--(6.053,3.693);
\draw[gp path] (6.173,0.691)--(6.173,0.781);
\draw[gp path] (6.173,3.783)--(6.173,3.693);
\draw[gp path] (6.277,0.691)--(6.277,0.781);
\draw[gp path] (6.277,3.783)--(6.277,3.693);
\draw[gp path] (6.369,0.691)--(6.369,0.781);
\draw[gp path] (6.369,3.783)--(6.369,3.693);
\draw[gp path] (6.452,0.691)--(6.452,0.871);
\draw[gp path] (6.452,3.783)--(6.452,3.603);
\node[gp node center] at (6.452,0.475) {$1000$};
\draw[gp path] (6.993,0.691)--(6.993,0.781);
\draw[gp path] (6.993,3.783)--(6.993,3.693);
\draw[gp path] (7.310,0.691)--(7.310,0.781);
\draw[gp path] (7.310,3.783)--(7.310,3.693);
\draw[gp path] (7.535,0.691)--(7.535,0.781);
\draw[gp path] (7.535,3.783)--(7.535,3.693);
\draw[gp path] (7.709,0.691)--(7.709,0.781);
\draw[gp path] (7.709,3.783)--(7.709,3.693);
\draw[gp path] (7.852,0.691)--(7.852,0.781);
\draw[gp path] (7.852,3.783)--(7.852,3.693);
\draw[gp path] (7.972,0.691)--(7.972,0.781);
\draw[gp path] (7.972,3.783)--(7.972,3.693);
\draw[gp path] (8.077,0.691)--(8.077,0.781);
\draw[gp path] (8.077,3.783)--(8.077,3.693);
\draw[gp path] (8.169,0.691)--(8.169,0.781);
\draw[gp path] (8.169,3.783)--(8.169,3.693);
\draw[gp path] (8.251,0.691)--(8.251,0.871);
\draw[gp path] (8.251,3.783)--(8.251,3.603);
\node[gp node center] at (8.251,0.475) {$10000$};
\draw[gp path] (8.251,0.691)--(8.071,0.691);
\node[gp node left] at (8.380,0.691) {$0.1$};
\draw[gp path] (8.251,1.001)--(8.161,1.001);
\draw[gp path] (8.251,1.183)--(8.161,1.183);
\draw[gp path] (8.251,1.312)--(8.161,1.312);
\draw[gp path] (8.251,1.411)--(8.161,1.411);
\draw[gp path] (8.251,1.493)--(8.161,1.493);
\draw[gp path] (8.251,1.562)--(8.161,1.562);
\draw[gp path] (8.251,1.622)--(8.161,1.622);
\draw[gp path] (8.251,1.675)--(8.161,1.675);
\draw[gp path] (8.251,1.722)--(8.071,1.722);
\node[gp node left] at (8.380,1.722) {$1$};
\draw[gp path] (8.251,2.032)--(8.161,2.032);
\draw[gp path] (8.251,2.213)--(8.161,2.213);
\draw[gp path] (8.251,2.342)--(8.161,2.342);
\draw[gp path] (8.251,2.442)--(8.161,2.442);
\draw[gp path] (8.251,2.524)--(8.161,2.524);
\draw[gp path] (8.251,2.593)--(8.161,2.593);
\draw[gp path] (8.251,2.652)--(8.161,2.652);
\draw[gp path] (8.251,2.705)--(8.161,2.705);
\draw[gp path] (8.251,2.752)--(8.071,2.752);
\node[gp node left] at (8.380,2.752) {$10$};
\draw[gp path] (8.251,3.063)--(8.161,3.063);
\draw[gp path] (8.251,3.244)--(8.161,3.244);
\draw[gp path] (8.251,3.373)--(8.161,3.373);
\draw[gp path] (8.251,3.473)--(8.161,3.473);
\draw[gp path] (8.251,3.554)--(8.161,3.554);
\draw[gp path] (8.251,3.623)--(8.161,3.623);
\draw[gp path] (8.251,3.683)--(8.161,3.683);
\draw[gp path] (8.251,3.736)--(8.161,3.736);
\draw[gp path] (8.251,3.783)--(8.071,3.783);
\node[gp node left] at (8.380,3.783) {$100$};
\draw[gp path] (1.054,3.783)--(1.054,0.691)--(8.251,0.691)--(8.251,3.783)--cycle;
\node[gp node center,rotate=-270] at (0.462,2.237) {time (s)};
\node[gp node center,rotate=-270] at (8.875,2.237) {mem. usage (GB)};
\node[gp node center] at (4.652,0.151) {number of tasks N};
\node[gp node right] at (3.479,3.490) {Parameterized (time)};
\gpcolor{rgb color={0.000,0.376,0.678}}
\gpsetlinewidth{2.00}
\draw[gp path] (3.608,3.490)--(4.304,3.490);
\draw[gp path] (1.054,1.493)--(1.596,1.505)--(2.312,1.536)--(2.853,1.584)--(4.111,1.835)%
  --(4.653,2.006)--(5.910,2.485)--(6.452,2.720)--(6.993,2.988)--(7.709,3.354)--(8.251,3.678);
\gpsetpointsize{4.00}
\gppoint{gp mark 7}{(1.054,1.493)}
\gppoint{gp mark 7}{(1.596,1.505)}
\gppoint{gp mark 7}{(2.312,1.536)}
\gppoint{gp mark 7}{(2.853,1.584)}
\gppoint{gp mark 7}{(4.111,1.835)}
\gppoint{gp mark 7}{(4.653,2.006)}
\gppoint{gp mark 7}{(5.910,2.485)}
\gppoint{gp mark 7}{(6.452,2.720)}
\gppoint{gp mark 7}{(6.993,2.988)}
\gppoint{gp mark 7}{(7.709,3.354)}
\gppoint{gp mark 7}{(8.251,3.678)}
\gppoint{gp mark 7}{(3.956,3.490)}
\gpcolor{color=gp lt color border}
\node[gp node right] at (3.479,3.265) {Parameterized (mem)};
\gpcolor{rgb color={0.000,0.565,0.867}}
\draw[gp path] (3.608,3.265)--(4.304,3.265);
\draw[gp path] (1.054,0.966)--(1.596,0.969)--(2.312,0.981)--(2.853,1.010)--(4.111,1.210)%
  --(4.653,1.375)--(5.910,1.927)--(6.452,2.175)--(6.993,2.473)--(7.709,2.851)--(8.251,3.160);
\gppoint{gp mark 5}{(1.054,0.966)}
\gppoint{gp mark 5}{(1.596,0.969)}
\gppoint{gp mark 5}{(2.312,0.981)}
\gppoint{gp mark 5}{(2.853,1.010)}
\gppoint{gp mark 5}{(4.111,1.210)}
\gppoint{gp mark 5}{(4.653,1.375)}
\gppoint{gp mark 5}{(5.910,1.927)}
\gppoint{gp mark 5}{(6.452,2.175)}
\gppoint{gp mark 5}{(6.993,2.473)}
\gppoint{gp mark 5}{(7.709,2.851)}
\gppoint{gp mark 5}{(8.251,3.160)}
\gppoint{gp mark 5}{(3.956,3.265)}
\gpcolor{color=gp lt color border}
\node[gp node right] at (3.479,3.040) {In-context (time)};
\gpcolor{rgb color={0.867,0.094,0.122}}
\draw[gp path] (3.608,3.040)--(4.304,3.040);
\draw[gp path] (1.054,1.358)--(1.596,1.389)--(2.312,1.429)--(2.853,1.514)--(4.111,1.944)%
  --(4.653,2.234)--(5.910,3.127)--(6.452,3.597);
\gppoint{gp mark 4}{(1.054,1.358)}
\gppoint{gp mark 4}{(1.596,1.389)}
\gppoint{gp mark 4}{(2.312,1.429)}
\gppoint{gp mark 4}{(2.853,1.514)}
\gppoint{gp mark 4}{(4.111,1.944)}
\gppoint{gp mark 4}{(4.653,2.234)}
\gppoint{gp mark 4}{(5.910,3.127)}
\gppoint{gp mark 4}{(6.452,3.597)}
\gppoint{gp mark 4}{(3.956,3.040)}
\gpcolor{color=gp lt color border}
\node[gp node right] at (3.479,2.815) {In-context (mem)};
\gpcolor{rgb color={0.992,0.220,0.247}}
\draw[gp path] (3.608,2.815)--(4.304,2.815);
\draw[gp path] (1.054,0.951)--(1.596,0.952)--(2.312,0.961)--(2.853,0.972)--(4.111,1.236)%
  --(4.653,1.461)--(5.910,2.373)--(6.452,2.876);
\gppoint{gp mark 1}{(1.054,0.951)}
\gppoint{gp mark 1}{(1.596,0.952)}
\gppoint{gp mark 1}{(2.312,0.961)}
\gppoint{gp mark 1}{(2.853,0.972)}
\gppoint{gp mark 1}{(4.111,1.236)}
\gppoint{gp mark 1}{(4.653,1.461)}
\gppoint{gp mark 1}{(5.910,2.373)}
\gppoint{gp mark 1}{(6.452,2.876)}
\gppoint{gp mark 1}{(3.956,2.815)}
\gpcolor{color=gp lt color border}
\gpsetlinewidth{1.00}
\draw[gp path] (1.054,3.783)--(1.054,0.691)--(8.251,0.691)--(8.251,3.783)--cycle;
%% coordinates of the plot area
\gpdefrectangularnode{gp plot 1}{\pgfpoint{1.054cm}{0.691cm}}{\pgfpoint{8.251cm}{3.783cm}}
\end{tikzpicture}
%% gnuplot variables